\documentclass[prb,twocolumn,superscriptaddress]{revtex4} 
\usepackage[colorlinks=true,linkcolor=blue,citecolor=blue]{hyperref}
\usepackage{graphicx}
\usepackage{amssymb}
\usepackage{amsmath}
\begin{document}
\title{\Large Theory and simulation of the diffusion of kinks on dislocations in bcc metals}
\author{T D Swinburne}
\email{tds110@ic.ac.uk}
\affiliation{Department of Physics, Imperial College London, Exhibition Road, London SW7 2AZ, UK}
\affiliation{EURATOM/CCFE Fusion Association, Culham Centre for Fusion Energy, Abingdon, Oxfordshire OX14 3DB, UK}
\author{S L Dudarev}
\affiliation{EURATOM/CCFE Fusion Association, Culham Centre for Fusion Energy, Abingdon, Oxfordshire OX14 3DB, UK}
\author{S P Fitzgerald}
\affiliation{EURATOM/CCFE Fusion Association, Culham Centre for Fusion Energy, Abingdon, Oxfordshire OX14 3DB, UK}
\author{M R Gilbert}
\affiliation{EURATOM/CCFE Fusion Association, Culham Centre for Fusion Energy, Abingdon, Oxfordshire OX14 3DB, UK}
\author{A P Sutton}
\affiliation{Department of Physics, Imperial College London, Exhibition Road, London SW7 2AZ, UK}
\date{\today}

\begin{abstract}
Isolated kinks on thermally fluctuating $(1/2)\langle111\rangle$ screw, $\langle 100\rangle$ edge and $(1/2)\langle111\rangle$ edge dislocations in bcc iron are simulated under zero stress conditions using molecular dynamics (MD). Kinks are seen to perform stochastic motion in a potential landscape that depends on the dislocation character and geometry, and their motion provides fresh insight into the coupling of dislocations to a heat bath. The kink formation energy, migration barrier and friction parameter are deduced from the simulations. A discrete Frenkel-Kontorova-Langevin (FKL) model is able to reproduce the coarse grained data from MD at $\sim$$10^{-7}$ of the computational cost, without assuming an \textit{a priori} temperature dependence beyond the fluctuation-dissipation theorem. Analytic results reveal that discreteness effects play an essential r\^ole in thermally activated dislocation glide, revealing the existence of a crucial intermediate length scale between molecular and dislocation dynamics. The model is used to investigate dislocation motion under the vanishingly small stress levels found in the evolution of dislocation microstructures in irradiated materials.
\end{abstract}

\maketitle
Dislocation motion is limited by two general processes, the formation and migration of kinks and pinning by impurities and other defects \cite{Argon}. In this paper we investigate the motion of kink-limited screw and edge dislocations in bcc Fe, where the kink formation energy is much larger than the thermal energy. To obtain dislocation motion on the time-scales accessible to molecular dynamics (MD) simulations some researchers have resorted to inducing kink formation by applying stresses some six orders of magnitude greater than those pertaining experimentally \cite{Gilbert2011,Terentyev2009}. But dislocation core structures and Peierls barriers are known to be highly stress-dependent\cite{Rodney2009}, making it difficult to relate simulation to the vanishingly low stress conditions found in thermally activated evolution of dislocation microstructures.\\

To avoid the problem of nucleating kinks in MD simulations we use periodic boundary conditions which enforce the existence of an isolated kink on each dislocation line in the supercell. {{This allows us to study the detailed dynamics of kink migration in isolation, a crucial and previously unexplored aspect of kink-limited dislocation motion, as seperate from the kink nucleation process.}} Under no applied stress, kinks are seen to undergo stochastic motion in a potential landscape that varies with the dislocation character and Burgers vector. A coarse graining procedure is introduced, that facilitates statistical analysis yielding many properties of the kink motion, and which provides physical insight into the coupling of dislocations to the heat bath. In particular, the friction parameter for a dislocation is found to be temperature-independent, contradicting decades of theoretical work since Liebfried\cite{Leibfried}. But this temperature-independence is seen in many other investigations of dislocations\cite{Gilbert,Gilbert2011,Terentyev2009} with a large lattice resistance and in the stochastic motion of point defects\cite{Dudarev2008b,Marian2002}, although to the best of our knowledge it has not been highlighted before.\\

Our main result is that a stochastic, discrete line representation of the dislocation, the discrete Frenkel-Kontorova-Langevin (FKL) model\cite{Braun1998}, is able to reproduce quantitatively the coarse grained data obtained from MD simulations of thermally activated dislocation glide at $\sim$$10^{-7}$ of the computational cost, with no \textit{a priori} assumption of any temperature dependence beyond the fluctuation-dissipation theorem. The computational efficiencies of the model are exploited to investigate dislocation motion under experimental stress levels inaccessible to MD.\\

We find the discreteness of the model, which is determined by the underlying crystallography, is essential to produce the thermally activated dynamics of dislocations seen in atomistic simulation. This shows that in order to model the kinetics of thermally activated dislocation motion beyond the limitations of atomistic simulation, a coarse grained model has to be sensitive to length scales smaller than those considered in traditional dislocation dynamics simulations.\\

The paper is structured as follows. In section \ref{sec:MD} we briefly review dislocation glide in bcc metals and the kink mechanism. We present results from large scale MD simulations of isolated kinks on $(1/2)\langle111\rangle$ screw, $(1/2)\langle111\rangle$ edge and $\langle100\rangle$ edge dislocation dipoles in bcc Fe. Kinks are given a crystallographic label, the kink vector, which enables a systematic enumeration of the kinks a dislocation line may support. The observed asymmetry between left and right kinks on screw dislocations is rationalized in terms of the kink vector, with the kink `core' being essentially symmetric. At finite temperature, under zero applied stress, kinks are seen in the simulations to perform stochastic motion in a potential landscape that varies with the dislocation character and geometry. We then introduce our coarse graining procedure.\\

In section \ref{sec:LE} we review the FKL model, using analytic expressions for kink properties to determine the parameters of the FKL model from MD simulations. The FKL model is known to lack any interaction between kinks \cite{Seeger}; we show that this deficiency has a very small effect on the kink formation energy and may be neglected in the investigation of very well separated kinks, allowing us to deduce parameters of the FKL model for the host dislocation. Numerical integration of the stochastic equations of motion of the FKL model produces data which may be processed identically to that from atomistic simulation, allowing us to compare the statistical results obtained from both methods. We find the transport properties of kinks in the MD simulations and their FKL counterparts to be in excellent agreement over a wide range of temperature for different dislocations. The parametrized FKL models are used to investigate screw and edge dislocation mobilities at applied stresses too low to induce dislocation motion in MD.\\
\section{Molecular dynamics simulation of thermally fluctuating dislocation lines}\label{sec:MD}
The periodic potential in the slip plane of a crystal leads to stable positions for a straight dislocation line separated by maxima in the potential energy known as Peierls barriers.  When the Peierls barrier is large compared to the available thermal energy, the mobility of dislocations is limited by the discreteness of the crystal structure. In that case dislocation glide takes place by the thermally activated nucleation and propagation of kinks\cite{Hirth}, which are localized regions connecting dislocation segments lying in adjacent valleys of the potential in the slip plane. The existence of kinks is clearly exhibited in MD simulations reported here and in many other investigations\cite{Terentyev2009,BulatovBook,Dudarev2011}, and their movement effects the glide of the dislocations on which they lie. To understand the kink mechanism it is necessary to investigate both kink propagation and nucleation of pairs of kinks. \\

When the formation energy of kink pairs is large compared to the thermal energy it becomes impossible to obtain statistically significant data on kink nucleation within MD time scales without resorting to unrealistic applied stresses, typically six orders of magnitude larger than those encountered in experiments\cite{Gilbert2011}. But dislocation core structures and Peierls barriers are known to change with applied stress \cite{Rodney2009}, making it difficult to relate such simulations to experimental reality. As a result, while the kink mechanism is well established in dislocation theory there is a sparsity of MD data on the parameters controlling kink motion, without which it is impossible to predict the velocity-stress relationship of the host dislocation for realistic stresses.\\

To circumvent the problem of kink nucleation in MD we use boundary conditions on the simulation supercell which enforce the existence of an isolated kink on the dislocation line. We use the MD code LAMMPS\cite{LAMMPS} with a recently developed potential by Gordon \textit{et al.}\cite{NewMendelev} which gives the best available representation of the screw dislocation core structure and bulk phonon dispersion. To avoid free surfaces, periodic boundary conditions must be imposed. The dislocation supercell must contain defects with no net Burgers vector to avoid a divergent elastic energy, and in this work we use dislocation dipoles. Thus the supercell contains two dislocations with equal and opposite Burgers vectors, and each dislocation has one kink in the supercell. \\

Dislocation dipoles were introduced by removing an appropriate number of atoms for an edge dislocation dipole or shearing the simulation supercell for a screw dislocation dipole\cite{DFTScrew}, and then applying the anisotropic elastic displacement field for the dipole. The system was relaxed by a conjugate gradients algorithm, followed by an annealing process which heated the system to 200K then back to zero temperature over 100ps (100,000 timesteps) to ensure that the system was in the ground state. To heat the system, atomic velocities were gradually rescaled according to a Maxwell-Boltzmann distribution of increasing temperature. Once the desired temperature was achieved the system was  evolved microcanonically and data was taken. This has a firmer statistical basis than using a thermostat because it relies on the real atomistic heat bath of a large system rather than any particular thermostat algorithm. However, unrealistic results can be obtained if there is significant heat generation or absorption as this may affect the probability of other activated processes\cite{Gilbert2011}. This is likely to be the case, for example, when a high energy kink pair annihilates or is created during the simulation. However, due to their large activation energies such processes did not occur during the simulations with the zero stress conditions investigated here, and the system temperature was observed to be constant throughout the simulation runs.\\

\begin{figure}[ht!]
\includegraphics[width=0.8\linewidth]{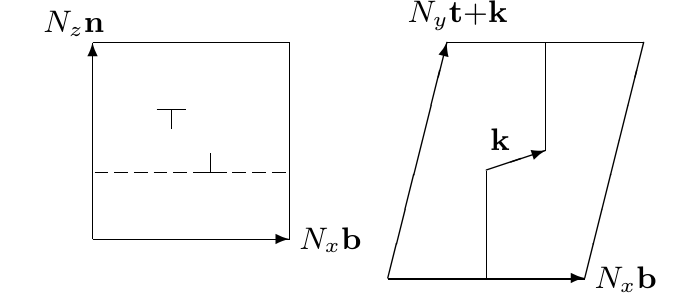}
\caption{Cartoon plan of the simulation supercell for an edge dislocation dipole, formed from lattice vectors $N_x\bf{b}$,  $N_y\bf{t}+\bf{k}$ and $N_z\bf{n}$. The broken line in the left figure highlights one of the two slip planes, which are separated by half the supercell height ${1\over2}N_z|\bf{n}|$.\label{101}}
\end{figure}

A kink connects two straight dislocation segments both parallel to a lattice vector $\bf{t}$ lying in the same slip plane. For the segments to be crystallographically equivalent they must be separated by a lattice vector, which may be uniquely identified, modulo $\bf t$, with a lattice vector $\bf k$ which we call the `kink vector'. Some dislocations may exist with a variety of core structures and there is a corresponding variety of atomic structures of kinks \cite{Bulatov97,Wang2003}. Nevertheless, the kink vector uniquely identifies any kink on a dislocation line in a given slip plane with a given Burgers vector and the same core structure on either side of the kink. We note in passing that a similar classification may also be applied to jogs, the sessile equivalent of a kink\cite{Hirth}, where the `jog vector' will be a lattice vector with a component normal to the slip plane. While the term kink vector has already appeared in the literature\cite{Wang2003,Ventelon2009} there has been no attempt to relate it to the host crystal lattice. To see the utility of our definition, consider a relaxed straight dislocation dipole, parallel to a lattice vector $\bf{t}$, in a supercell formed from lattice vectors $N_x\bf{m}$, $N_y\bf{t}$ and $N_z\bf{n}$, where the $N_{i=x,y,z}$ are all integers. To impose the boundary conditions required for a kink on each dislocation we create a new supercell from $N_x\bf{m}$,  $N_y\bf{t}+\bf{k}$ and $N_z\bf{n}$, as shown in Figure \ref{101}. In principle $\bf k$ may be any lattice vector modulo $\bf t$ lying in the slip plane, but for high index lattice vectors the relaxed structure will possess multiple kinks as one might expect if $\bf k$ spans many minima of the potential in the slip plane. The set of low index lattice vectors lying in the slip plane quickly provides an enumeration of the possible kinks a straight dislocation line may support, generalizing the approach taken in previous work\cite{Ventelon2009} to produce isolated kinks on dislocation lines. We have used this procedure to generate the simulation supercell geometries employed here. Each supercell contained approximately 700,000 atoms, with the dislocation line either initially sharply kinked as in Figure \ref{101} or parallel to the $N_y\bf{t}+\bf{k}$ supercell vector. The relaxed configurations were independent of this initial preparation.\\

\begin{figure}
\includegraphics[width=\linewidth]{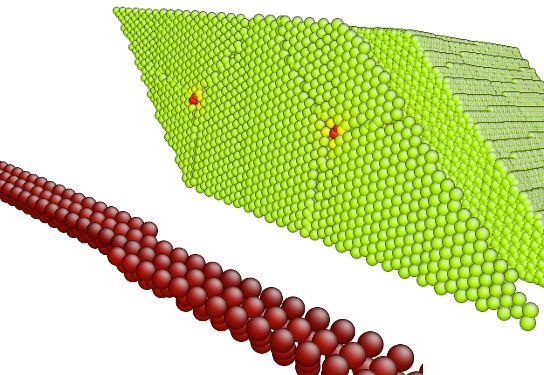}
\caption{Above: An atomic plane normal to a screw dislocation dipole is exposed with the higher energy atoms colored progressively red. The dislocation cores are clearly identified. Below: Plotting only the highest energy atoms in successive atomic planes (red) along the supercell reveals a kinked dislocation line. (Color online) \label{slice} }
\end{figure}
\subsection{Coarse graining procedure for atomistic simulation}
Having obtained a relaxed configuration, the atoms were grouped into atomic planes normal to the unkinked dislocation line direction $\bf{t}$. The potential energy in each plane has clearly defined peaks, the centres of gravity of which identify the positions of the dislocation core, as illustrated in Figure \ref{slice}. The coarse grained representation of the dislocation comprises a line threading nodes at each of these core positions. There is one node for each atomic plane normal to the dislocation line. The coarse grained representation is well defined, independent of dislocation character, and it yields the position of the dislocation with atomic resolution. The unique mapping between the atomistic and coarse grained representations enables the energy of the coarse grained representation to be determined at each time step of the simulation. The position of each node on the dislocation line moves in one dimension, normal to $\bf t$ and the slip plane normal $\bf n$. The kink position and width were determined from the center and width of the maximum in the core energy along the dislocation line, with the trajectories of the two kinks in each supercell each forming a time series $\{x_{n\Delta{t}}\}$,$n=0,1,...N$. The kink position is readily located graphically as shown in Figure \ref{110_kink}, where the dislocation segments on either side of the kink are straight lines at absolute zero. The dynamics of the kink positions are simulated with MD in the following section and we aim to reproduce the dynamics with a coarse grained, many body, stochastic model.\\

The effectiveness of the coarse graining relies on the uniformity of the potential energy of atoms in the bulk; however, at finite temperature it is expected and observed that random fluctuations in atomic positions and energies due to thermal vibrations obscure the dislocation position. To filter out this noise it is necessary to average atomic positions and energies over a period of a few thermal oscillations. The atomic coordinates and energies can then be processed in an identical fashion to the relaxed zero temperature system, again yielding localized dislocation core positions. Kinks appear as localized geometric and energetic regions along the coarse grained dislocation line, allowing the determination of the kink position, width and formation energy by calculating the total deviation from the core energy of a straight dislocation line. \\

Other techniques to determine the dislocation position are to calculate the deviation of the atomic displacements from the anisotropic elastic field\cite{Ventelon2009}, or the bonding disregistry across the slip plane\cite{Clouet2008}. However, we found that the procedure employed here gave better localization at finite temperature and is applicable to many different dislocation geometries.\\
\subsection{Stochastic motion of isolated kinks on edge dislocations}
Table \ref{kvs} shows kink formation energies calculated for edge dislocations in bcc Fe. These values were obtained by calculating the excess energy in a cylindrical slice coaxial with the average dislocation line direction, relative to the energy of a slice containing the same number of atoms for a straight dislocation. The slices contain one atomic plane normal to the dislocation lines. The radius of the cylinder was enlarged, using the periodic boundary conditions if necessary to generate atomic coordinates outside the supercell, until the excess energy of the dislocation core in the slice reached a constant asymptotic value. This excess core energy is plotted in green in Figure \ref{110_kink}. The kink formation energy is the sum of these excess core energies along the dislocation line. Convergence in the core energy per atomic plane was typically achieved for a supercell length of thirty Burgers vectors for kinks on edge dislocations. This implies that the interaction energy between a kink and its periodic images along the line is not detectable at separations of more than thirty Burgers vectors. The kink formation energies are in broad agreement with other studies\cite{LiQun,Terentyev2009,Chang}.

\begin{table}[!h]
\begin{small}
\begin{tabular*}{0.48\textwidth}{@{\extracolsep{\fill}}ccccc}\toprule
Burgers & Glide & Tangent & Kink & Formation \\
vector ($\bf{b}$) &plane &vector ($\bf{t}$)& vector ($\bf{k}$) & energy \\
\colrule
 ${1\over2}[111]$ & $(1\bar{2}1)$ & $[10\bar{1}]$ & ${1\over2}[111]$ 	   & 0.15 eV\\
  		  & $(\bar{1}01)$ & $[1\bar{2}1]$ & ${1\over2}[1\bar{1}1]$ & 0.03 eV\\
  		  & 		  & 		  & ${1\over2}[1\bar{3}1]$ & 0.02 eV\\
	$[100]$   & 	$(001)$   & 	$[010]$   & $[100]$		   & 0.51 eV\\\
		   & $(011)$& $[01\bar{1}]$ & ${1\over2}[11\bar{1}]$ & 0.25 eV\\\botrule\\
\end{tabular*}
\caption{Kink vectors and fully relaxed kink formation energies (to 2 s.f.) on edge dislocations at absolute zero, calculated using the potential developed by Gordon \textit{et al.}\cite{NewMendelev} for bcc Fe\label{vectors}. The tangent vector $\bf{t}$ is the primitive lattice vector along the unkinked dislocation line. \label{kvs}}
\end{small}
\end{table}
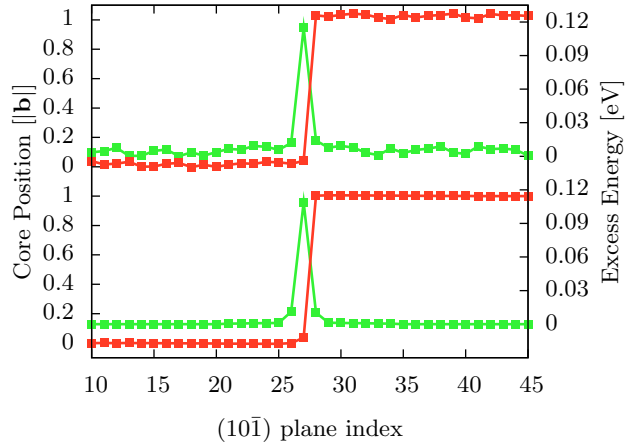
\begin{figure}
 \setlength{\unitlength}{0.0500bp}%
  \begin{picture}(4874.00,3628.00)%
      \put(528,815){\makebox(0,0)[r]{\strut{}0}}%
      \put(528,1036){\makebox(0,0)[r]{\strut{}0.2}}%
      \put(528,1258){\makebox(0,0)[r]{\strut{}0.4}}%
      \put(528,1480){\makebox(0,0)[r]{\strut{}0.6}}%
      \put(528,1701){\makebox(0,0)[r]{\strut{}0.8}}%
      \put(528,1923){\makebox(0,0)[r]{\strut{}1}}%
      \put(528,2144){\makebox(0,0)[r]{\strut{}0}}%
      \put(528,2366){\makebox(0,0)[r]{\strut{}0.2}}%
      \put(528,2587){\makebox(0,0)[r]{\strut{}0.4}}%
      \put(528,2809){\makebox(0,0)[r]{\strut{}0.6}}%
      \put(528,3031){\makebox(0,0)[r]{\strut{}0.8}}%
      \put(528,3252){\makebox(0,0)[r]{\strut{}1}}%
      \put(660,484){\makebox(0,0){\strut{} 10}}%
      \put(1130,484){\makebox(0,0){\strut{} 15}}%
      \put(1600,484){\makebox(0,0){\strut{} 20}}%
      \put(2070,484){\makebox(0,0){\strut{} 25}}%
      \put(2539,484){\makebox(0,0){\strut{} 30}}%
      \put(3009,484){\makebox(0,0){\strut{} 35}}%
      \put(3479,484){\makebox(0,0){\strut{} 40}}%
      \put(3949,484){\makebox(0,0){\strut{} 45}}%
      \put(4081,957){\makebox(0,0)[l]{\strut{}0}}%
      \put(4081,1210){\makebox(0,0)[l]{\strut{}0.03}}%
      \put(4081,1464){\makebox(0,0)[l]{\strut{}0.06}}%
      \put(4081,1717){\makebox(0,0)[l]{\strut{}0.09}}%
      \put(4081,1970){\makebox(0,0)[l]{\strut{}0.12}}%
      \put(4081,2223){\makebox(0,0)[l]{\strut{}0}}%
      \put(4081,2477){\makebox(0,0)[l]{\strut{}0.03}}%
      \put(4081,2730){\makebox(0,0)[l]{\strut{}0.06}}%
      \put(4081,2983){\makebox(0,0)[l]{\strut{}0.09}}%
      \put(4081,3236){\makebox(0,0)[l]{\strut{}0.12}}%
      \put(154,2033){\rotatebox{-270}{\makebox(0,0){\strut{}Core Position [$|{\bf b}|$]}}}%
      \put(4586,2033){\rotatebox{-270}{\makebox(0,0){\strut{}Excess Energy [eV]}}}%
      \put(2304,154){\makebox(0,0){\strut{}$(10\bar{1})$ plane index}}%
    \put(0,0){\includegraphics{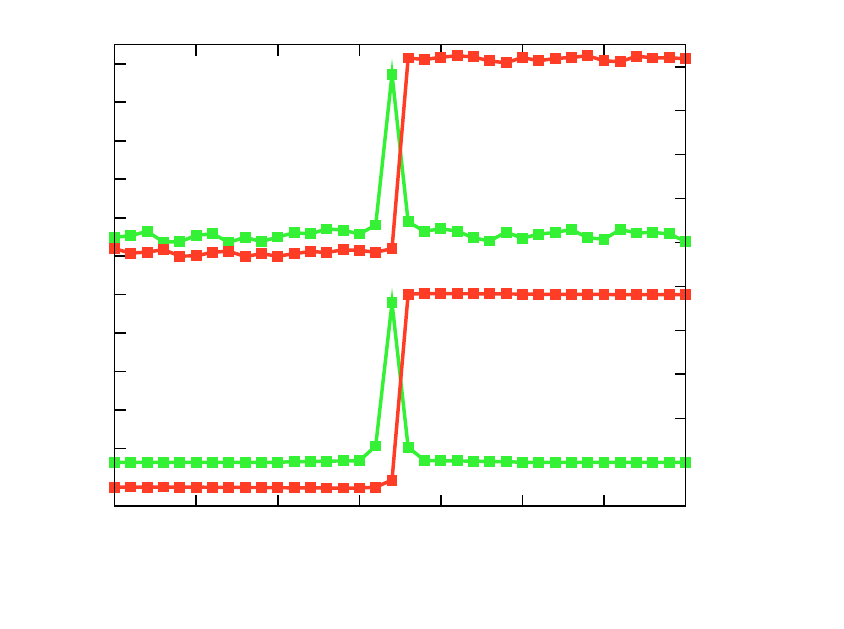}}%
  \end{picture}%
\caption{Illustration of the coarse grained data from atomistic simulation of a kink on a $(1/2)[111](1\bar{2}1)$ edge dislocation line at 0K (bottom) and 90K (top). The kink is clearly localized as measured by the position of the core (red) and the core energy (green). Note the narrow kink width in contrast to the screw dislocation kinks in Figure \ref{screw_E}. (Color online) \label{110_kink}}
\end{figure}

As $[100](011)$ edge dislocations have not been directly observed in experiment they are of little interest and we do not consider them further here. 

The very low formation energy of 0.03 eV for kinks on $(1/2)\langle 111\rangle\{1\bar{1}0\}$ edge dislocations indicates that the mobility of these dislocations is not limited by kinks except possibly at the very lowest temperatures. Therefore we investigate isolated kinks on $(1/2)[111](1\bar{2}1)$ and $[100](010)$ edge dislocations, whose motion is known to be kink-limited, at temperatures up to which kinks remain isolated on MD time-scales.\\

\begin{figure}
\includegraphics[width=0.4\textwidth]{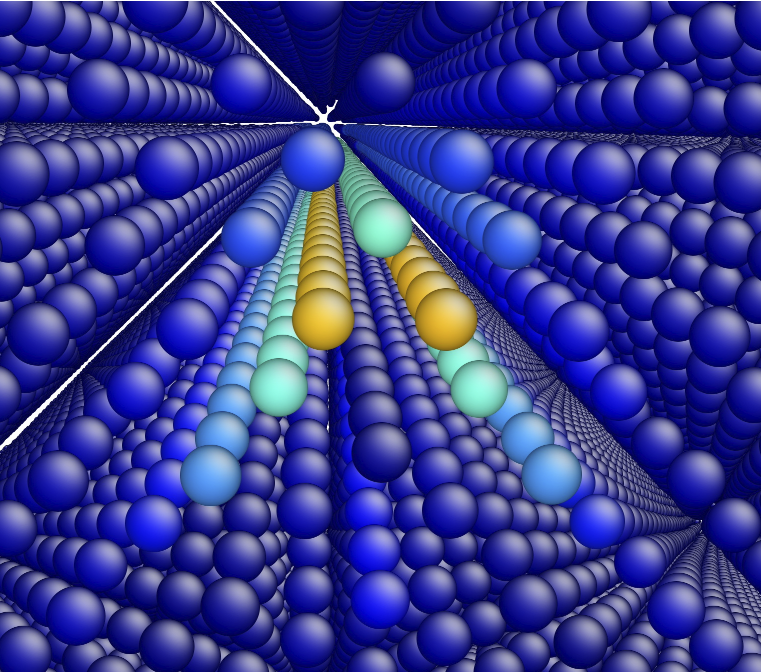}\\
\vspace{0.2cm}
\includegraphics[width=0.35\textwidth]{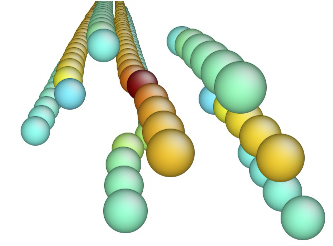}
\caption{Above: a $[100](010)$ edge dislocation in the bulk. Below: a kink on this dislocation with the bulk atoms removed. The non-planar core gives a large kink formation energy of 0.61eV. (Color Online) \label{100_kink}}
\end{figure}

The kink formation energy for ${1\over2}[111](1\bar{2}1)$ dislocations is 0.15eV. At temperatures below 300K it is possible to observe and analyze the stochastic motion of an isolated kink for MD runs of several nanoseconds. Similarly, for $[100](010)$ edge dislocations  no additional kinks are expected to be nucleated in MD runs of several nanoseconds at temperatures up to 700K owing to their large formation energy of 0.61eV. This large formation energy is due to the non-planar core, shown in Figure \ref{100_kink}. An isolated kink is localized geometrically and energetically as shown in Figure \ref{110_kink}. \\

\begin{figure}
  \setlength{\unitlength}{0.0500bp}%
  \begin{picture}(4874.00,3004.00)%
      \put(528,704){\makebox(0,0)[r]{\strut{}-100}}%
      \put(528,958){\makebox(0,0)[r]{\strut{} 0}}%
      \put(528,1213){\makebox(0,0)[r]{\strut{} 100}}%
      \put(528,1467){\makebox(0,0)[r]{\strut{} 200}}%
      \put(528,1722){\makebox(0,0)[r]{\strut{} 300}}%
      \put(528,1976){\makebox(0,0)[r]{\strut{} 400}}%
      \put(528,2230){\makebox(0,0)[r]{\strut{} 500}}%
      \put(528,2485){\makebox(0,0)[r]{\strut{} 600}}%
      \put(528,2739){\makebox(0,0)[r]{\strut{} 700}}%
      \put(660,484){\makebox(0,0){\strut{} 0}}%
      \put(1208,484){\makebox(0,0){\strut{} 200}}%
      \put(1756,484){\makebox(0,0){\strut{} 400}}%
      \put(2305,484){\makebox(0,0){\strut{} 600}}%
      \put(2853,484){\makebox(0,0){\strut{} 800}}%
      \put(3401,484){\makebox(0,0){\strut{} 1000}}%
      \put(3949,484){\makebox(0,0){\strut{} 1200}}%
      \put(22,1721){\rotatebox{-270}{\makebox(0,0){\strut{}Kink Displacement [${\rm\AA}$]}}}%
      \put(2304,154){\makebox(0,0){\strut{}Time [ps]}}%
      \put(2005,884){\makebox(0,0)[l]{\strut{}T=90K}}%
      \put(3358,2566){\makebox(0,0)[r]{\strut{}Kink 1}}%
      \put(3358,2346){\makebox(0,0)[r]{\strut{}Kink 2}}%
    \put(0,0){\includegraphics{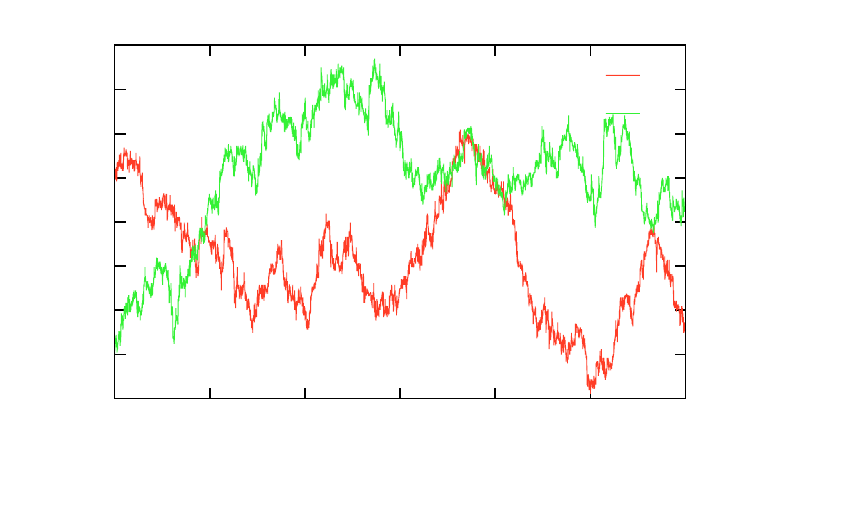}}%
    \end{picture}%

\caption{The trajectories of the two kinks on a ${1\over 2}[111](1\bar{2}1)$ edge dislocation dipole at T=90K. (Color online)\label{traj}}
\end{figure}
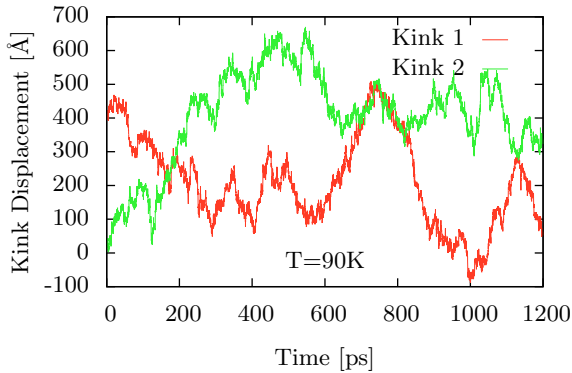

\begin{figure}
\setlength{\unitlength}{0.0500bp}%
  \begin{picture}(4874.00,3004.00)%
      \put(528,704){\makebox(0,0)[r]{\strut{}-3}}%
      \put(528,995){\makebox(0,0)[r]{\strut{}-2}}%
      \put(528,1285){\makebox(0,0)[r]{\strut{}-1}}%
      \put(528,1576){\makebox(0,0)[r]{\strut{} 0}}%
      \put(528,1867){\makebox(0,0)[r]{\strut{} 1}}%
      \put(528,2158){\makebox(0,0)[r]{\strut{} 2}}%
      \put(528,2448){\makebox(0,0)[r]{\strut{} 3}}%
      \put(528,2739){\makebox(0,0)[r]{\strut{} 4}}%
      \put(910,484){\makebox(0,0){\strut{} 0.005}}%
      \put(1460,484){\makebox(0,0){\strut{} 0.006}}%
      \put(2010,484){\makebox(0,0){\strut{} 0.007}}%
      \put(2560,484){\makebox(0,0){\strut{} 0.008}}%
      \put(3110,484){\makebox(0,0){\strut{} 0.009}}%
      \put(3660,484){\makebox(0,0){\strut{} 0.01}}%
      \put(286,1721){\rotatebox{-270}{\makebox(0,0){\strut{}$\ln\left(\bar{D}\right)$}}}%
      \put(2304,154){\makebox(0,0){\strut{}1/T [K$^{-1}$]}}%
      \put(895,1275){\makebox(0,0)[l]{\strut{}$E_{mig}$ = 74meV}}%
      \put(895,1072){\makebox(0,0)[l]{\strut{}$\gamma_{kink}$ = 1.79 m$_u$ps$^{-1}$}}%
      \put(3358,2566){\makebox(0,0)[r]{\strut{}Diffusion Constant $\bar{D}$}}%
      \put(3358,2346){\makebox(0,0)[r]{\strut{}Equation (\ref{BessD})}}%
    \put(0,0){\includegraphics{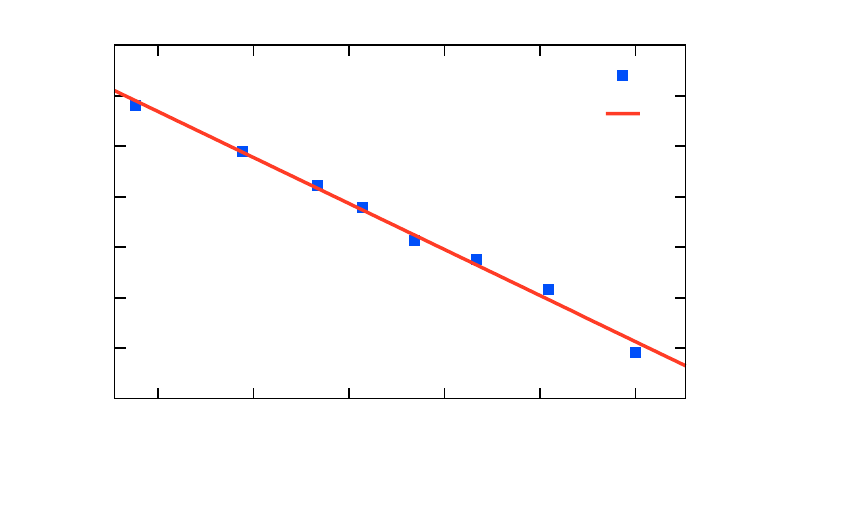}}%
  \end{picture}%
\caption{Arrhenius plot of the diffusion constant $\bar{D}$ for the kink center of mass on a $(1/2)[111](1\bar{2}1)$ edge dislocation dipole. A two-parameter fit of equation (\ref{BessD}) gives a migration barrier of 74meV, {a large fraction of the 150meV kink formation energy.} The linear gradient implies the dissipation parameter $\gamma_{kink}$ is independent of temperature (see text). (Color Online) \label{Arrhenius}}
\end{figure}
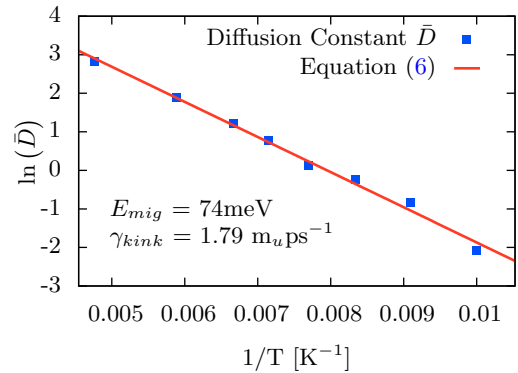

Figure \ref{traj} shows the trajectories of the two kinks in a simulation supercell on a $(1/2)\langle111\rangle(1\bar{1}0)$ edge dislocation dipole. The two kinks appear to be moving independently, but we cannot be sure that there is no significant interaction between them.  Any correlation arising from their interaction may be eliminated by analyzing the center of mass $\bar{x}$, defined here to be the mean of the two kink positions, $\bar{x}=(x^{(1)}+x^{(2)})/2$. It may be shown\cite{Dudarev2010,Reichl2009} that such a quantity is independent of any interaction, and it yields a diffusion constant one half that of a free kink, $D_{kink}/2$. Thus we construct from the two kink positions a single time series $\{\bar{x}_{n\Delta{t}}\},n=0,1,...N$ for the center of mass to ensure such correlation effects do not affect our results. We look for diffusive behavior in the mean squared displacement (MSD) $\langle{\Delta\bar{x}^2}\rangle$ over a range of intervals $\tau$, defined as
\begin{eqnarray}
\langle{\Delta\bar{x}^2}\rangle(\tau) &=& \sum_{n=0}^{N-\tau/\Delta{t}}{(\bar{x}_{n\Delta{t}+\tau}-\bar{x}_{n\Delta{t}})^2\over{N-\tau/\Delta{t}}}\\\nonumber
&&-\left(\sum_{n=0}^{N-\tau/\Delta{t}}{(\bar{x}_{n\Delta{t}+\tau}-x_{n\Delta{t}})\over{N-\tau/\Delta{t}}}\right)^2,\label{MDMSD}
\end{eqnarray}
which is the variance of the displacement. It is well known\cite{Reichl2009} that for diffusive motion with a diffusion constant $D_{kink}/2$,
\begin{equation}
\langle{\Delta\bar{x}^2}\rangle(\tau) = D_{kink}\tau.\label{MDMSD_D}
\end{equation}\\
\begin{figure}
  \setlength{\unitlength}{0.0500bp}%
  \begin{picture}(4704.00,3004.00)%
      \put(528,704){\makebox(0,0)[r]{\strut{} 0}}%
      \put(528,1043){\makebox(0,0)[r]{\strut{} 50}}%
      \put(528,1382){\makebox(0,0)[r]{\strut{} 100}}%
      \put(528,1722){\makebox(0,0)[r]{\strut{} 150}}%
      \put(528,2061){\makebox(0,0)[r]{\strut{} 200}}%
      \put(528,2400){\makebox(0,0)[r]{\strut{} 250}}%
      \put(528,2739){\makebox(0,0)[r]{\strut{} 300}}%
      \put(660,484){\makebox(0,0){\strut{} 5}}%
      \put(1106,484){\makebox(0,0){\strut{} 10}}%
      \put(1551,484){\makebox(0,0){\strut{} 15}}%
      \put(1997,484){\makebox(0,0){\strut{} 20}}%
      \put(2442,484){\makebox(0,0){\strut{} 25}}%
      \put(2888,484){\makebox(0,0){\strut{} 30}}%
      \put(3333,484){\makebox(0,0){\strut{} 35}}%
      \put(3779,484){\makebox(0,0){\strut{} 40}}%
      \put(22,1721){\rotatebox{-270}{\makebox(0,0){\strut{}$\langle \bar{x}^2\rangle$ [${\rm\AA}^2$]}}}%
      \put(2219,154){\makebox(0,0){\strut{}Time Interval $T=M\Delta$ [ps]}}%
      \put(1251,2566){\makebox(0,0)[l]{\strut{}T=100K}}%
      \put(1251,2346){\makebox(0,0)[l]{\strut{}T=130K}}%
      \put(1251,2126){\makebox(0,0)[l]{\strut{}T=170K}}%
      \put(1251,1906){\makebox(0,0)[l]{\strut{}Linear Fit}}%
    \put(0,0){\includegraphics{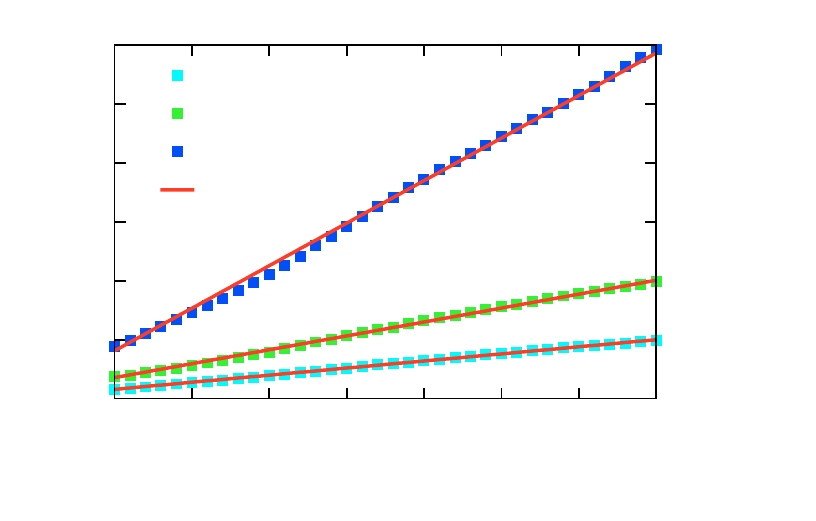}}%
  \end{picture}%
\caption{The mean square displacement as defined in equation (\ref{MDMSD}) for the kink center of mass on a $(1/2)[111](1\bar{2}1)$ edge dislocation dipole. The linear relationship with time is in agreement with diffusive behavior, equation (\ref{MDMSD_D}).\label{OUedge}}
\end{figure}
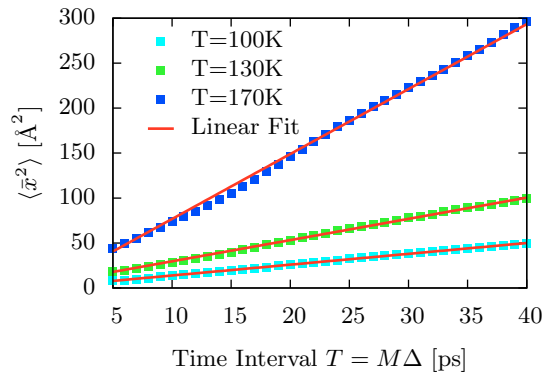
Examples of the MSD are shown for the system of kinks on edge dislocations considered above in Figure \ref{OUedge}. The MSD clearly shows the linear time dependence characteristic of diffusive behavior, with the diffusion constant as defined in equation (\ref{MDMSD_D}) shown in Figure \ref{Arrhenius} for kinks on ${1\over 2}[111](1\bar{2}1)$ edge dislocations. Kinks on $a[100](010)$ edge dislocations exhibit similar behavior. The diffusivity rises exponentially with temperature in both cases, indicating that the kink motion is thermally activated across the kink migration barrier\cite{Hirth}. We therefore conclude that the kink performs one dimensional stochastic motion in a periodic migration potential $V(x+a)$=$V(x)$ whose amplitude $E_{mig}$=$V_{MAX}-V_{MIN}$ is large compared to the thermal energy. While the traditional analysis for such data is to fit an Arrhenius form $D_0\exp(-E_{mig}/k_BT)$ for the diffusion constant, in one dimension there exists an exact solution, given by the Lifson-Jackson formula\cite{Lifson} 
\begin{equation}
D_{kink} = {k_BTa^2\over\gamma_{kink}}\left(\int_0^a{e^{-V(x)/k_BT} dx} \int_0^a{e^{V(x)/k_BT}dx}\right)^{-1}\label{LJ},
\end{equation}
where $k_B$ is Boltzmann's constant and $\gamma_{kink}$ is the friction, or dissipation, parameter\cite{coffey2004}, which measures the rate of momentum transfer from the diffusing object (here a kink) to the heat bath. $\gamma_{kink}$ plays a key r\^ole in the stochastic equations of motion introduced in section \ref{sec:LE}, defining the frictional force $-\gamma_{kink}v$ and it is the inverse of the kink mobility. To gain insight into equation (\ref{LJ}) we investigate limiting cases. When the amplitude of the migration potential $E_{mig}$=$V_{MAX}-V_{MIN}$ is much greater than thermal energy $k_BT$, as for the case of kinks on edge dislocations here, we may evaluate the integrals in (\ref{LJ}) by the method of steepest descents. Denoting $V''$ for the second derivative, (\ref{LJ}) becomes
\begin{equation}
D_{kink} \simeq a^2{\sqrt{V''_{MIN}V''_{MAX}}\over2\pi\gamma_{kink}}e^{-({V_{MAX}-V_{MIN})/ k_BT}}\label{kramersD},
\end{equation}
which is precisely the Arrhenius form given by Kramers\cite{Hanggi} for thermally activated diffusive motion. We note that the traditional temperature independence of the prefactor in (\ref{kramersD}) requires that $\gamma_{kink}$ be independent of temperature. In the other limit, when the thermal energy $k_BT$ is much larger than $E_{mig}$, the integrals (\ref{LJ}) are constant, giving a diffusivity 
\begin{equation}
D_{kink} \simeq {k_BT\over\gamma_{kink}}\label{freeD}
\end{equation}
as first described by Einstein\cite{einstein1956} for a freely diffusing particle. We note that a linear temperature dependence in the diffusivity (\ref{freeD}) implies  $\gamma_{kink}$ is again independent of temperature. We will see that kinks on screw dislocations exhibit the diffusive behavior of (\ref{freeD}) due to their negligible migration barrier and thus the relation (\ref{LJ}) is able to capture the wide range of diffusive behavior exhibited by kinks on dislocation lines. A numerical illustration of (\ref{LJ}) is shown in Figure \ref{TH}, where we indeed see the failure of the Arrhenius law when $k_BT \gg E_{mig}$. In section \ref{sec:LE} we show that to an excellent approximation the migration barrier for a kink is sinusoidal, $V(x) = E_{mig}\sin^2(\pi x/a)$, allowing an exact expression of (\ref{LJ})
\begin{equation}
D_{kink} ={k_BT\over\gamma_{kink}}{1\over I^{2}_0(E_{mig}/2k_BT)},
\label{BessD}
\end{equation}
where $I_0(x)$ is the zeroth order modified Bessel function\cite{Abram}. We thus perform a two parameter fit of (\ref{BessD}) to the kink diffusion constant with temperature to determine $\gamma_{kink}$ and $E_{mig}$ for the kink systems investigated here, the results of which are shown in Table \ref{Kink_Table}.\\
There are two points of note in the MD results for kinks on edge dislocations in Table \ref{Kink_Table}. Firstly, we find that the migration barrier is comparable to the formation energy, implying that the nature of the kink mechanism on edge dislocations is complex, with double kink nucleation and kink migration occurring on similar timescales. This agrees with previous simulations on edge dislocations\cite{Terentyev2009} where the mobility was found to be independent of the dislocation segment length. Secondly, the linear gradient of the Arrhenius plot in Figure \ref{Arrhenius} also implies, by equation (\ref{kramersD}), that the dissipation parameter for the kink $\gamma_{kink}$ is temperature independent. \\

We show in section \ref{sec:LE} that the dissipation parameter for a kink is proportional to the dissipation parameter for the host dislocation line, which should therefore also be temperature independent. {{This is in agreement with several other studies of dislocations with a large lattice resistance\cite{Gilbert}, self-interstitial defects\cite{Dudarev2008b} and prismatic loops\cite{Marian2002} in bcc Iron. However, decades of theoretical work\cite{Leibfried,Nabarro,Hirth} conclude that the dissipation parameter for a dislocation should increase linearly with temperature due to the increasing phonon population, as has been found by simulation for dislocations with a negligible lattice resistance, such as $(1/2)[111](1\bar{2}1)$ edge dislocations\cite{Gilbert2011} in bcc Iron and many dislocations in fcc metals\cite{RodneyBacon}.}} We return to this important issue concerning the coupling of dislocations to the heat bath in the following section on screw dislocations, as the diffusive form (\ref{freeD}) they exhibit allows an even more direct investigation of $\gamma_{kink}$.\\
\begin{figure}
  \setlength{\unitlength}{0.0500bp}%
  \begin{picture}(4874.00,3004.00)%
      \put(660,484){\makebox(0,0){\strut{} 0}}%
      \put(1154,484){\makebox(0,0){\strut{} 0.2}}%
      \put(1649,484){\makebox(0,0){\strut{} 0.4}}%
      \put(2143,484){\makebox(0,0){\strut{} 0.6}}%
      \put(2637,484){\makebox(0,0){\strut{} 0.8}}%
      \put(3132,484){\makebox(0,0){\strut{} 1}}%
      \put(3626,484){\makebox(0,0){\strut{} 1.2}}%
      \put(4120,484){\makebox(0,0){\strut{} 1.4}}%
      \put(440,1721){\rotatebox{-270}{\makebox(0,0){\strut{}D [Arbitrary Units]}}}%
      \put(2436,154){\makebox(0,0){\strut{}kT [Barrier Height]}}%
      \put(1251,2566){\makebox(0,0)[l]{\strut{}Lifson-Jackson Form, Eq. (\ref{BessD})}}%
      \put(1251,2346){\makebox(0,0)[l]{\strut{}Arrhenius Form}}%
      \put(1251,2126){\makebox(0,0)[l]{\strut{}Stochastic Simulation}}%
    \put(0,0){\includegraphics{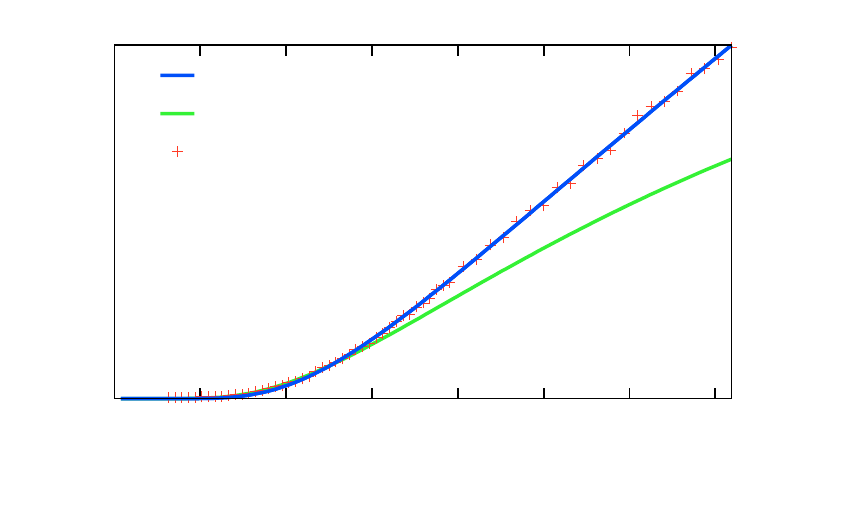}}%
  \end{picture}%
\caption{Diffusivity in a one dimensional periodic potential. Equation (\ref{BessD}) (blue), the appropriate Arrhenius form (green) and numerical data (red) are compared across a wide temperature range. At low temperatures all three agree but at intermediate to high temperatures a linear temperature dependence emerges in simulation and equation (\ref{BessD}). (Color Online)\label{TH}}
\end{figure}
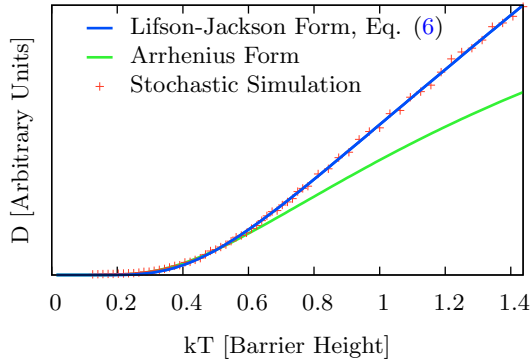\subsection{Stochastic motion of isolated kinks on screw dislocation lines} 
While edge dislocations may be thought of as an inserted half plane of atoms\cite{Hirth} producing a bonding disregistry perpendicular to the dislocation line direction, screw dislocations create a bonding disregistry along the dislocation line direction, which does not require the addition or removal of material. Screw dislocations possess a non planar core structure in bcc metals, which gives a large Peierls barrier\cite{Vitek1974,Vitek2004}. The complex core structure is heavily influenced by the choice of interatomic potential used in classical atomistic simulations. The vast majority of existing potentials predict a screw dislocation has multiple core structures\cite{Dudarev,Mendelev}, leading many authors to suggest that a screw dislocation may pass through a metastable core structure during the kink nucleation process\cite{Vitek2004}. Under an applied stress this can produce a new kink formation pathway leading to a discontinuity in the flow stress\cite{Argon,NewMendelev,Gilbert}. However, this discontinuity is not shown in experiment, and recent \textit{ab initio} calculations\cite{JAP,VentDFT} rule out any metastable core structure, with the nucleation pathway seen to occur almost entirely in the $\{1\bar{1}0\}$ slip planes. A recently developed potential by Gordon \textit{et al.}\cite{NewMendelev} attempts to address these issues,  concluding that while the metastable core may be removed from the nucleation pathway, multiple core structures remain. Using this potential, we introduce kinks with the core structure predicted from first principles calculations, thereby minimizing unphysical effects due to the interatomic potential.\\

A screw dislocation dipole requires a triclinic simulation cell to avoid spurious image stresses; we refer the reader elsewhere for details of the simulation method\cite{Cai2004,BulatovBook,Clouet2011} which are well established. As before, the kink vector was added to the supercell vectors to give the boundary conditions required for isolated kinks to form on each dislocation under relaxation. {{As there is no mirror symmetry along $\langle 111\rangle$ directions in the bcc lattice, the so called `right' and `left' kinks forming a kink pair on a screw dislocation will have different kink vectors and thus are expected to be asymmetric.}} Previous zero temperature calculations of isolated kinks on screw dislocations in bcc Iron\cite{Ventelon2009} with the Mendelev \textit{et al.} potential\cite{Mendelev} found a noticeable difference between the formation energies of right and left kinks, which correspond to kink vectors ${\bf k}_R$=$(1/2)[1\bar{1}1]$ and ${\bf k}_L$=$[010]$. They also found the kink formation energy converged to a constant value when the supercell length was greater than the widths of two kinks. The supercell length measures the separation between a kink and its periodic images. These findings are inconsistent with elasticity theory of kink interactions\cite{Seeger}, according to which the far field interaction between kinks should decrease with the inverse of the kink separation.\\

To investigate these discrepancies we performed similar calculations with the improved potential by Gordon \textit{et al.}, extending the supercell length to 240 Burgers vectors, more than double that used in \cite{Ventelon2009}. In agreement with elasticity theory we found that when the supercell length was greater than two kink widths the kink formation energy decreased with the inverse of the supercell length, by 0.011eV for both right and left kinks over a distance of 200 Burgers vectors. The formation energy for the right and left kinks was 0.604eV and 0.13eV respectively, giving a double kink formation energy of 0.734eV, in good agreement with the previous study.\\

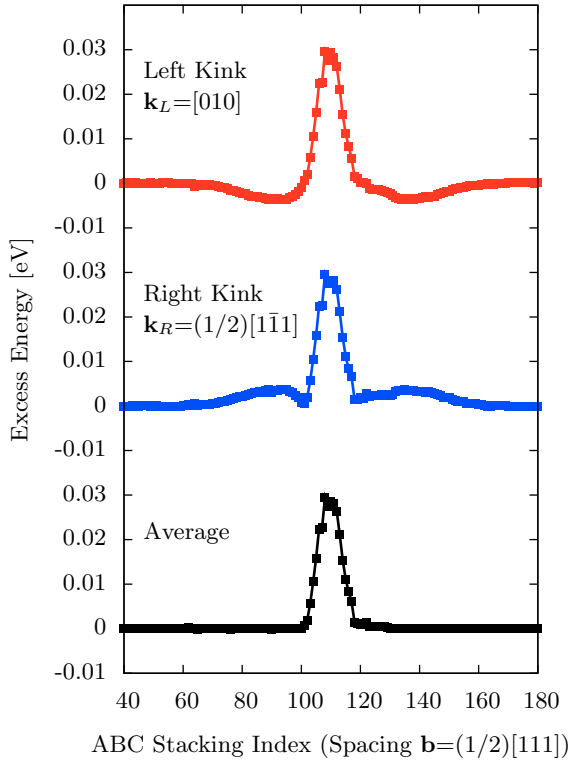
\begin{figure}
\setlength{\unitlength}{0.0500bp}%
  \begin{picture}(4704.00,6008.00)%
      \put(528,704){\makebox(0,0)[r]{\strut{}-0.01}}%
      \put(528,1040){\makebox(0,0)[r]{\strut{}0}}%
      \put(528,1376){\makebox(0,0)[r]{\strut{}0.01}}%
      \put(528,1712){\makebox(0,0)[r]{\strut{}0.02}}%
      \put(528,2048){\makebox(0,0)[r]{\strut{}0.03}}%
      \put(528,2384){\makebox(0,0)[r]{\strut{}-0.01}}%
      \put(528,2720){\makebox(0,0)[r]{\strut{}0}}%
      \put(528,3056){\makebox(0,0)[r]{\strut{}0.01}}%
      \put(528,3391){\makebox(0,0)[r]{\strut{}0.02}}%
      \put(528,3727){\makebox(0,0)[r]{\strut{}0.03}}%
      \put(528,4063){\makebox(0,0)[r]{\strut{}-0.01}}%
      \put(528,4399){\makebox(0,0)[r]{\strut{}0}}%
      \put(528,4735){\makebox(0,0)[r]{\strut{}0.01}}%
      \put(528,5071){\makebox(0,0)[r]{\strut{}0.02}}%
      \put(528,5407){\makebox(0,0)[r]{\strut{}0.03}}%
      \put(660,484){\makebox(0,0){\strut{} 40}}%
      \put(1106,484){\makebox(0,0){\strut{} 60}}%
      \put(1551,484){\makebox(0,0){\strut{} 80}}%
      \put(1997,484){\makebox(0,0){\strut{} 100}}%
      \put(2442,484){\makebox(0,0){\strut{} 120}}%
      \put(2888,484){\makebox(0,0){\strut{} 140}}%
      \put(3333,484){\makebox(0,0){\strut{} 160}}%
      \put(3779,484){\makebox(0,0){\strut{} 180}}%
      \put(806,5236){\makebox(0,0)[l]{\strut{}Left Kink}}%
      \put(806,5001){\makebox(0,0)[l]{\strut{}${\bf k}_L$=$[010]$}}%
      \put(806,3556){\makebox(0,0)[l]{\strut{}Right Kink}}%
      \put(806,3321){\makebox(0,0)[l]{\strut{}${\bf k}_R$=$(1/2)[1\bar{1}1]$}}%
      \put(806,1756){\makebox(0,0)[l]{\strut{}Average}}%
      \put(-110,3223){\rotatebox{-270}{\makebox(0,0){\strut{}Excess Energy [eV]}}}%
      \put(2219,154){\makebox(0,0){ABC Stacking Index (Spacing $\bf b$=$(1/2)[111]$)}}%
    \put(0,0){\includegraphics{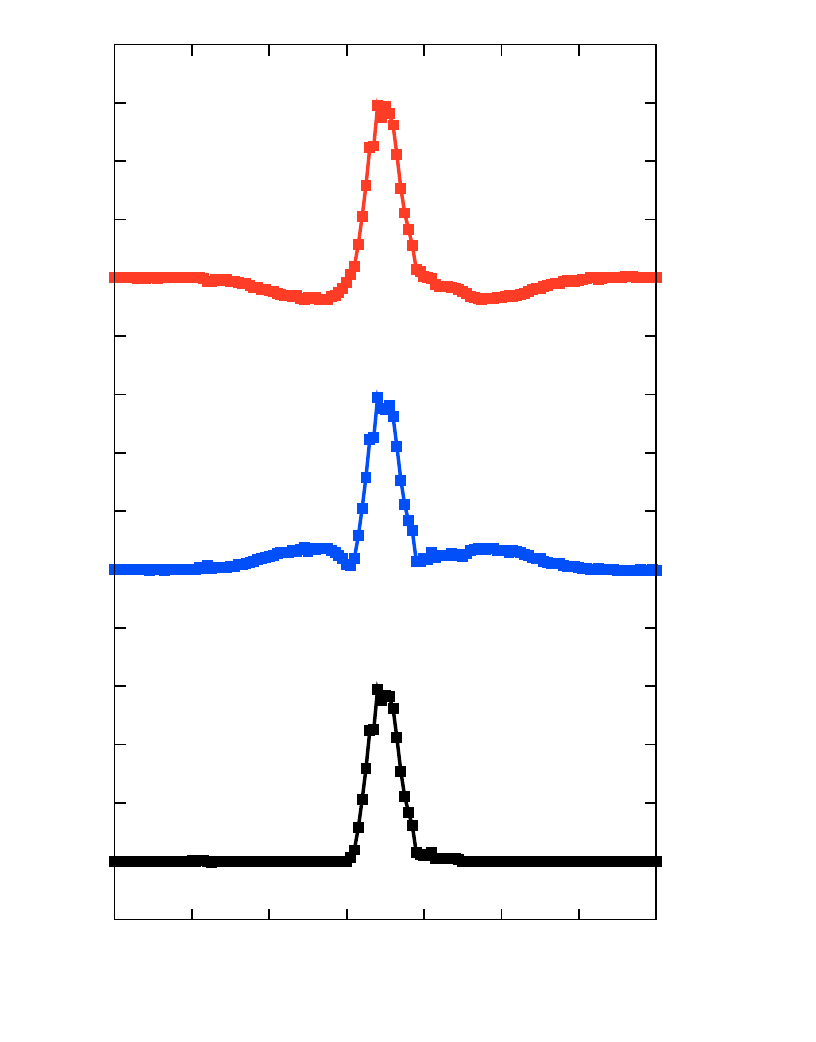}}%
    \end{picture}%
\caption{Excess energy per ABC stacking sequence (see text) for right (blue) and left (red) kinks on a $(1/2)[111](1\bar{1}0)$ screw dislocation. The kinks appear as  peaks of similar height with asymmetric tails. The tails are removed under averaging (black) as described in the text. Note the large kink width as compared to the edge dislocation kinks in Figure \ref{110_kink}. (Color Online)\label{screw_E}}
\vspace{0.15cm}
\end{figure}

 We conclude that the long range kink interaction, while decaying inversely with separation as predicted by elasticity theory\cite{Seeger}, is a small perturbation to the kink formation energy in the atomistic simulations performed here and in \cite{Ventelon2009}. However, the difference in  formation energies of left and right kinks is still unexplained. To gain insight into the kink structure, Figure \ref{screw_E} shows the excess potential energy, relative to a straight dislocation, per atomic plane normal to the line of a screw dislocation with a right or left kink, obtained by the coarse graining procedure described above. The kinks appear as well defined peaks of approximately the same height but also with long range tails which differ markedly between the two kinks. These long range tails are the source of the difference in the formation energies, with the `core' of each kink very similar in size and energy.\\
{{The nature of these tails can be understood by noting that the kinks may be regarded as two short segments of dislocation with edge character. However, the edge segments are of equal and opposite sign because their line senses are reversed. For the right kink we see a projection of the kink vector along the dislocation line of $|{\bf b}|/3$, which implies the insertion of two atomic planes and is often thought of as an interstitial kink\cite{Ventelon2009}, whose compressive far field locally raises the core energy of the host dislocation. For the left kink we see a projection of the kink vector along the dislocation line of $2|{\bf b}|/3$, which is most appropriately thought of as the removal of $6-4$=$2$ planes due to the stacking sequence of the dislocation core, and is often referred to as a vacancy kink\cite{Ventelon2009}. The tensile far field then lowers the core energy locally due to a slight relaxation towards the bulk crystal structure. This picture also implies that the kink fields should cancel at long range.}} To test this we average the energy per unit length for the left and right kinked screw dislocation lines, shown in Figure \ref{screw_E}, where we indeed see a localized peak with a width of around 20$|{\bf b}|$, much wider than the sharp kinks of width 3$|{\bf b}|$ seen on edge dislocations. It is this core energy and width which we take to define kinks on screw dislocations.\\

\begin{figure}
\setlength{\unitlength}{0.0500bp}%
  \begin{picture}(4874.00,3004.00)%
      \put(528,704){\makebox(0,0)[r]{\strut{} 0}}%
      \put(528,930){\makebox(0,0)[r]{\strut{} 20}}%
      \put(528,1156){\makebox(0,0)[r]{\strut{} 40}}%
      \put(528,1382){\makebox(0,0)[r]{\strut{} 60}}%
      \put(528,1608){\makebox(0,0)[r]{\strut{} 80}}%
      \put(528,1835){\makebox(0,0)[r]{\strut{} 100}}%
      \put(528,2061){\makebox(0,0)[r]{\strut{} 120}}%
      \put(528,2287){\makebox(0,0)[r]{\strut{} 140}}%
      \put(528,2513){\makebox(0,0)[r]{\strut{} 160}}%
      \put(528,2739){\makebox(0,0)[r]{\strut{} 180}}%
      \put(660,484){\makebox(0,0){\strut{} 50}}%
      \put(1130,484){\makebox(0,0){\strut{} 100}}%
      \put(1600,484){\makebox(0,0){\strut{} 150}}%
      \put(2070,484){\makebox(0,0){\strut{} 200}}%
      \put(2539,484){\makebox(0,0){\strut{} 250}}%
      \put(3009,484){\makebox(0,0){\strut{} 300}}%
      \put(3479,484){\makebox(0,0){\strut{} 350}}%
      \put(3949,484){\makebox(0,0){\strut{} 400}}%
      \put(22,1721){\rotatebox{-270}{\makebox(0,0){\strut{}$D_{kink}$ [${\rm\AA}^2ps^{-1}$]}}}%
      \put(2304,154){\makebox(0,0){\strut{}T [K]}}%
      \put(1251,2566){\makebox(0,0)[l]{\strut{}Left $\gamma_{kink}=1.98$ m$_u$ps$^{-1}$}}%
      \put(1251,2346){\makebox(0,0)[l]{\strut{}Right $\gamma_{kink}=1.69$ m$_u$ps$^{-1}$}}%
      \put(1251,2126){\makebox(0,0)[l]{\strut{}Linear Fit}}%
      \put(851,1906){\makebox(0,0)[l]{\strut{}($D_{kink}=T/\gamma_{kink}$)}}%
    \put(0,0){\includegraphics{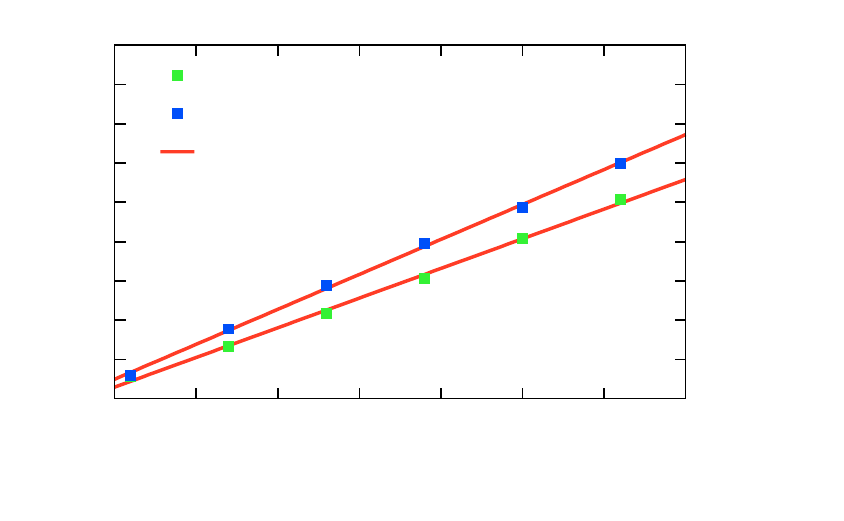}}%
  \end{picture}%
\caption{Diffusion constant for left and right kinks. A linear temperature dependence is exhibited, which implies the dissipation parameter does not depend on temperature. (Color Online)\label{screw_D}}
\vspace{0.15cm}
\end{figure}
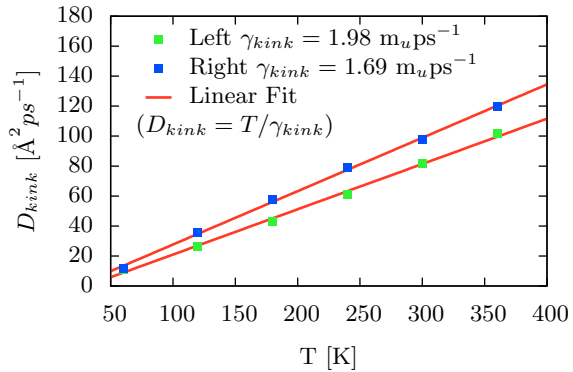

At finite temperature, the kink trajectories were analyzed in a similar manner to that detailed above for edge dislocations. However, the temperature dependence of the kink diffusivities, shown in Figure \ref{screw_D}, is markedly different. They exhibit a linear temperature dependence, which by equations (\ref{LJ}), (\ref{freeD}) implies a negligible migration barrier, as found in static calculations\cite{Ventelon2009}. It also implies that the dissipation parameter $\gamma_{kink}$ is independent of temperature. \\

The temperature independence is significant as all theories of dislocation damping since Liebfried\cite{Leibfried,Nabarro} have concluded that the dissipation parameter for a dislocation must increase linearly with temperature due to the increased phonon population. We emphasize that the temperature independence of the dissipation parameter is exhibited in both thermally activated diffusion, where the prefactor of the Arrhenius law is independent of temperature, and in essentially free diffusion, where the gradient of the diffusion constant with temperature is independent of temperature. {{The prediction of a linear temperature dependence in the dissipation parameter essentially arises due to the vanishingly small phase space predicted for one phonon scattering, meaning the leading contribution to the thermal force is a two phonon scattering term\cite{Dudarev2002}. A consequence of the fluctuation-dissipation theorem is that the variance of the thermal phonon force acting on a body is equal to the system temperature multiplied by the dissipation parameter for that body. It may be shown that the thermal force produced by one phonon scattering has a variance which increases linearly with temperature, meaning the dissipation parameter is temperature independent as seen here, whilst the thermal force produced by two phonon scattering has a variance which increases quadratically with temperature, giving a dissipation parameter which increase linearly with temperature, as predicted by the standard theory and as observed in the drift motion of dislocations with a low lattice resistance. We suggest that the presence of a strong lattice coupling leads to a strong enhancement of the one body phonon scattering term, giving a temperature independent dissipation of momentum, though it is clear that this discrepancy between theory and simulation remains unexplained and is an important topic for future investigation.}}

In this section we have reported the results from large scale molecular dynamics simulations of isolated kinks on edge and screw dislocation lines in bcc iron. The large simulation cells required significant computational power to obtain statistically significant results; however the total real time simulated was still of the order of nanoseconds. We now introduce a model which aims to reproduce the coarse grained data from the full atomistic simulation at a fraction of the computational cost, allowing access to experimentally relevant time and length scales.
\section{A stochastic model for a dislocation line}\label{sec:LE}
There are two principal methods for simulating dislocation motion that avoid an explicit treatment of atomic dynamics: dislocation dynamics and kinetic Monte Carlo methods. In conventional dislocation dynamics codes the motion of dislocations is entirely deterministic \cite{BulatovBook}; the stochastic dislocation dynamics observed in MD simulations and experimentally cannot be simulated with such codes. The traditional technique to model thermally dominated motion is a master equation approach\cite{Hudson2004,Kumar2012,Hudson2005}. This assigns probabilities from the canonical ensemble to transitions between different system states, which are then implemented in a kinetic Monte Carlo simulation. However, the large state space available to even an isolated flexible dislocation quickly renders the technique extremely cumbersome. The assignment of a canonical distribution is hard or impossible to justify in non-equilibrium environments and while the logarithmic time scale employed improves efficiency it obscures comparison to the real-time trajectories given by experiment and atomistic simulation.\\

In this section we introduce a model which aims to reproduce the results from full atomistic simulation. The model we employ is the well known discrete Frenkel-Kontorova-Langevin (FKL) model\cite{Braun1998,Kosevich2006,Joos1997} which treats the dislocation line as a discrete elastic string sitting in a periodic substrate potential. The representation of a dislocation as an elastic line was first used to model pinning by trapping sites\cite{Granato1956} and due to its simplicity equivalent systems appear in many areas\cite{Argon}. A discrete Langevin equation approach has recently been used to model the thermal motion of $(1/2)\langle{111}\rangle$ prismatic and vacancy dislocation loops in bcc Fe by Derlet \textit{et al.}\cite{Dudarev2011}. However, the absence of a substrate potential rendered the model unable to capture any kink mechanisms and the discreteness had no relation to the crystallography of the corresponding atomic system. In contrast, the spacing of nodes of the elastic string in our model is determined by the spacing of atomic planes normal to the dislocation line. We find this is essential to reproduce the structure and dynamics of kinks seen in atomistic simulations. While these discreteness effects have previously been investigated theoretically by Jo\'os and Duesbery\cite{Joos1997} in covalent materials there has been no investigation, to our knowledge, of the dynamical behavior they predict.

Many different shapes of the substrate potential used in the FKL model have been investigated\cite{Seeger}. However, the only qualitative change occurs in the presence of deep metastable minima, which imply the existence of a metastable core structure. As discussed above, even for the complex case of screw dislocations, recent \textit{ab initio} calculations show the kink formation process to take place in the slip plane, with no metastable core structure. This allows us to take the substrate potential in the FKL model as a simple sinusoid, and is consistent with the approximation of taking a dislocation line to be a string of constant internal structure moving only in the slip plane.\\

The main criticism of the FKL model is its inability to capture the long range kink interaction predicted from elasticity and the different formation energies of left and right kinks seen in atomistic simulation. However, we have seen that the long range interaction is a minor perturbation on the formation energy. Additionally, the asymmetric kink formation energy was seen to result from the long range fields of the kinks. The simplest term which captures this behavior is linear in the dislocation line gradient\cite{Braun1998} and consequently will not affect the equations of motion as it may be integrated out of the Lagrangian \cite{landau1975classical}. Investigation of more complicated terms is beyond the requirements of the current investigation as very little difference was found in the diffusivities of left and right kinks. In effect our model treats the localized, symmetric kink `cores' shown in the lower panel of Figure \ref{screw_E}.\\

First we obtain analytic expressions for the kink formation energy, width and migration barrier in terms of the parameters defining the FKL model. We then introduce the stochastic equations of motion which govern the system dynamics, obtaining an analytic expression identical in form to equation (\ref{LJ}) for the kink diffusivity. By equating these analytic expressions to the values obtained for the kink formation energy, width, migration barrier and diffusivity from MD, we may solve numerically for the FKL model parameters. The system is then stochastically integrated and compared to the output from full atomistic simulation.

\subsection{Static Properties}
The FKL model treats a dislocation line as a discrete elastic string sitting in a periodic substrate potential. The string is constructed from a set of harmonically coupled nodes spaced by a fixed distance $a$, which we set equal to the distance between atomic planes normal to the dislocation line. The string sits in a substrate potential of period $L_P$, which we set equal to the projection of the relevant kink vector normal to the dislocation line, often known as the kink height. As a result the two length scales of the model, $a$ and $L_P$, are determined by the crystallography of the corresponding atomistic system.\\

Taking a co-ordinate system $(x,y)$, where $\bf \hat x$ lies along the (unkinked) dislocation line direction and $\bf \hat y$ is normal to $\bf \hat x$ in the slip plane, each dislocation is represented by a discrete line of points $\{(na,u_n(t))\}$, where $n=0,1,2,..N$ and only the $\{u_n(t)\}$ vary with time. Each node is thus constrained to move only in the $\bf \hat y$ direction. The potential energy is as follows: 
\begin{equation}
V\left(\{u_n\}\right) = \sum_{n=0}^N aV_P\sin^2\left(\pi{u_n\over L_P}\right) + a{\kappa\over2}\left(u_{n+1}-u_{n} \over a\right)^2\label{DV},
\end{equation}
where $V_P$ is the amplitude of the substrate potential and $\kappa$ is the harmonic coupling strength, both in units of energy per unit length,
with displaced periodic boundary conditions to account for the presence of the kink
\begin{equation}
u_{n+N}(t) = u_n(t) + L_P.\label{KBC}
\end{equation}

\begin{figure}
\setlength{\unitlength}{0.0500bp}%
  \begin{picture}(4874.00,3004.00)%
      \put(528,704){\makebox(0,0)[r]{\strut{} 0}}%
      \put(528,1043){\makebox(0,0)[r]{\strut{} 0.1}}%
      \put(528,1382){\makebox(0,0)[r]{\strut{} 0.2}}%
      \put(528,1722){\makebox(0,0)[r]{\strut{} 0.3}}%
      \put(528,2061){\makebox(0,0)[r]{\strut{} 0.4}}%
      \put(528,2400){\makebox(0,0)[r]{\strut{} 0.5}}%
      \put(528,2739){\makebox(0,0)[r]{\strut{} 0.6}}%
      \put(660,484){\makebox(0,0){\strut{} 1}}%
      \put(1104,484){\makebox(0,0){\strut{} 1.5}}%
      \put(1548,484){\makebox(0,0){\strut{} 2}}%
      \put(1992,484){\makebox(0,0){\strut{} 2.5}}%
      \put(2437,484){\makebox(0,0){\strut{} 3}}%
      \put(2881,484){\makebox(0,0){\strut{} 3.5}}%
      \put(3325,484){\makebox(0,0){\strut{} 4}}%
      \put(3769,484){\makebox(0,0){\strut{} 4.5}}%
      \put(4213,484){\makebox(0,0){\strut{} 5}}%
      \put(40,1721){\rotatebox{-270}{\makebox(0,0){\strut{}Amplitude [$w_0V_P/4$]}}}%
      \put(2436,154){\makebox(0,0){\strut{}Kink Width $2w_0$ [$a$]}}%
      \put(3622,2566){\makebox(0,0)[r]{\strut{}$E_{mig} = \tilde{V}_{mig}(1)$}}%
      \put(3622,2306){\makebox(0,0)[r]{\strut{}$\tilde{V}_{mig}(2)$}}%
      \put(3622,2046){\makebox(0,0)[r]{\strut{}$\tilde{V}_{mig}(3)$}}%
    \put(0,0){\includegraphics{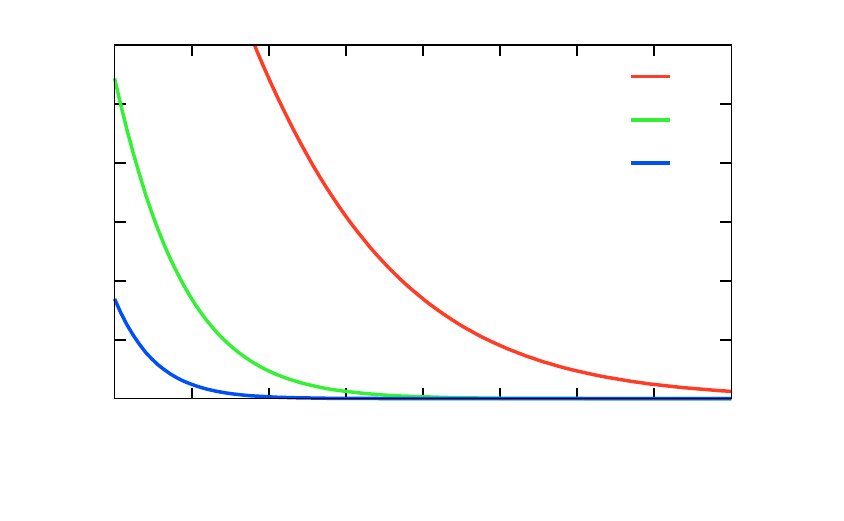}}%
  \end{picture}%
\caption{The magnitude of the first three summands in the kink energy (\ref{KMP}) as a function of the equilibrium kink width $w_0$. We see the leading term $\tilde{V}_{mig}(1)$ dominates but all three terms decrease rapidly with increasing $w_0$. (Color Online) \label{FKharm}}
\end{figure}
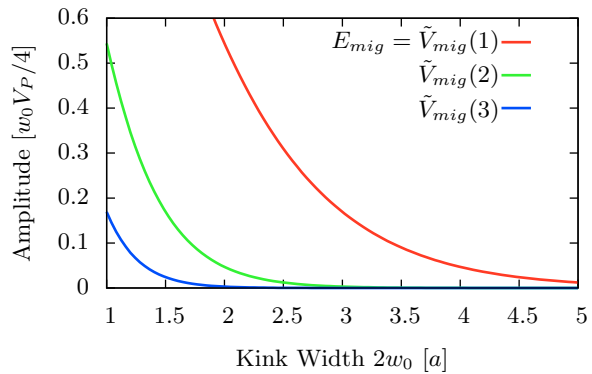

To obtain an analytic expression for the shape of the static kink we first take the continuum limit $a\rightarrow0, N\rightarrow\infty$. In this limit, the system energy (\ref{DV}) with boundary conditions (\ref{KBC}) is minimized by the soliton kink
\begin{equation}
u_{kink}(x-X)={L_P\over\pi}\left(\tan^{-1}\sinh\left(x-X\over w_0\right)+{\pi\over2}\right),\nonumber
\end{equation}
\begin{equation}
w_0 = {L_P\over2\pi}\sqrt{2\kappa\over V_P}\label{FKkink},
\end{equation}
where $X$ is the kink position and $2w_0$ is the kink width, which is proportional to $\sqrt{\kappa/V_P}$. An illustration of (\ref{FKkink}) is shown as the red curve in Figure \ref{kinkprof}. The soliton kink (\ref{FKkink}) shape interpolates the numerically minimized discrete structure of (\ref{DV}) well. However, in the continuum limit the system energy (\ref{DV}) is independent of the kink position, while in the discrete system the energy varies periodically with the kink position as the continuous translation symmetry is broken, in direct analogy to the Peierls barrier for a dislocation. This position dependent energy produces the kink migration barrier discussed above. It may be shown\cite{Kosevich2006,Joos1997,Seeger} that substituting (\ref{FKkink}) into (\ref{DV}) gives
\begin{equation}
V\left(\{u_{kink}(na-X)\}\right) = {w_0V_P\over 4} +\sum_{n=1}^\infty \tilde{V}_{mig}(n)\cos\left({2n\pi\over a}X\right),\nonumber
\end{equation}
\begin{equation}
\tilde{V}_{mig}(n)={V_P\over 4a}{n\pi w^2_0 \over\sinh\left(n\pi{w_0/a}\right)}.\label{KMP}
\end{equation}
The first few $\tilde{V}_{mig}(n)$ are shown as functions of the equilibrium kink width $2w_0$ in Figure \ref{FKharm}. We see that for realistic kink widths of more than $2a$ the leading term $\tilde{V}_{mig}(1)$ dominates by a factor of at least ten, allowing us to approximate the migration potential as a sinusoid of period $a$. The kink energy (\ref{KMP}) has a minimum when the kink center of mass lies between two nodes, $X = na+a/2$, as in this configuration no node lies at the maximum of the substrate potential. This minimum kink energy, which should be equated to the formation energy from the relaxed atomistic simulation, is thus
\begin{equation}
E_{kink} = {w_0V_P\over 4} - {E_{mig}\over 2}, \label{kink_energy}
\end{equation}
\begin{equation}
E_{mig} = 2\tilde{V}_{mig}(1) = {V_P\over 2a}{\pi w^2_0 \over\sinh\left(\pi{w_0/a}\right)}.\label{kink_mig}
\end{equation}
The form of the migration barrier $E_{mig}$ provides insight into the observed behavior of kinks in atomistic simulation. As can be seen in Figure \ref{FKharm}, $E_{mig}$ decreases rapidly with the equilibrium kink width, in agreement with the observation that the narrow kinks on edge dislocations have a significant migration barrier as compared to the essentially free motion of the wide kinks on screw dislocations. These significant effects would be entirely lost in any continuum model, emphasizing the importance of atomistic resolution in modeling dislocation dynamics.\\
\subsection{Dynamic Properties}
We now introduce the stochastic equations of motion which govern the dynamics of the FKL model (\ref{DV}). We recall that the atomistic data was time-averaged. Applying the same procedure to our model is formally equivalent to taking a strong damping limit\cite{coffey2004,Zwanzig} and thus permits a first order equation of motion for the node displacements. This approximation is supported by the absence of any ballistic motion, even over short time intervals, in the simulation data. The output from the stochastic model may thus be subjected to identical analysis as the output from atomistic simulation, allowing us to determine whether the data produced is statistically equivalent.\\
The thermal behavior of the system is investigated through the stochastic integration of first order Langevin equations\cite{coffey2004}, which balance a frictional force proportional to the velocity, $-\gamma_{line} v$, the conservative force $-\partial V / \partial u$ and a `fluctuation force' which will be detailed below. This simplified equation of motion allows us to integrate the system on a much coarser timescale. In addition, any notion of a dislocation mass is assigned to the dissipation parameter $\gamma_{line}$, which measures the rate of momentum transfer from the dislocation to the heat bath. In this way we avoid the controversial concept of dislocation inertia as inertial effects were not exhibited in the atomistic simulations we wish to reproduce with this model. The first order equation of motion for our discrete system (\ref{DV}), with boundary conditions (\ref{KBC}), is
\begin{equation}
\gamma_{line}{d u_n(t) \over dt}=-{\partial\over\partial u_n}V(\{u_m\}) + \eta_n(t) \label{lEoM}
\end{equation}
where $\gamma_{line}$ is the dissipation parameter for the dislocation line and the $\{\eta_n(t)\}$ are independent Gaussian random variables\cite{reif} representing the stochastic force from the surrounding heat bath. They are defined under an ensemble average $\langle . . .\rangle$ of all heat baths at a temperature $T$, possessing only an average and standard deviation by the central limit theorem. These read
\begin{equation}
\langle \eta_n(t)\rangle = 0,\quad\langle\eta_n(t)\eta_m(t')\rangle = 2\gamma_{line} T\delta_{nm}\delta(t'-t)\label{GWN}.
\end{equation}
The amplitude of the fluctuations $\sqrt{2\gamma_{line} T}$ is uniquely determined by the fluctuation-dissipation theorem, which requires that the steady state solution to the Fokker-Planck equation associated with (\ref{lEoM}) is the canonical distribution\cite{coffey2004}. The absence of any spatial correlation in the noise forces reflects the chaotic atomic dynamics of the surrounding heat bath and does not preclude any correlation in the dislocation motion; however the delta function $\delta(t'-t)$ is strictly the limiting case of a vanishingly small correlation time in the atomic collisions which constitute the heat bath\cite{reif}. This limit may be taken only when we operate on a sufficiently coarse timescale much longer than an individual collision, which is indeed the case in the first order equations of motion investigated here.\\

We now have a completely specified system; we will integrate the equations stochastically (\ref{lEoM}), using a pseudorandom number algorithm\cite{dSFMT} to generate the stochastic forces (\ref{GWN}). However, in order to have expressions for all the quantities extracted from atomistic simulation in terms of the model parameters we must also derive the kink diffusion constant.\\
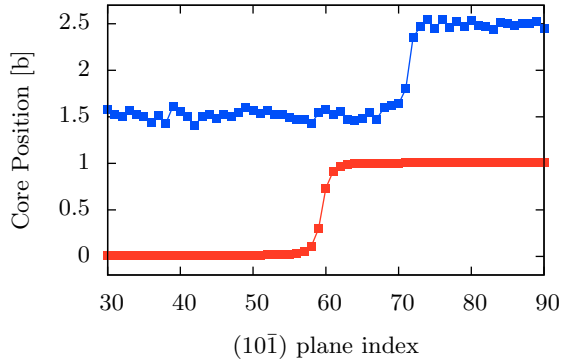
\begin{figure}
\setlength{\unitlength}{0.0500bp}%
  \begin{picture}(4874.00,3004.00)%
      \put(528,844){\makebox(0,0)[r]{\strut{} 0}}%
      \put(528,1195){\makebox(0,0)[r]{\strut{} 0.5}}%
      \put(528,1546){\makebox(0,0)[r]{\strut{} 1}}%
      \put(528,1897){\makebox(0,0)[r]{\strut{} 1.5}}%
      \put(528,2248){\makebox(0,0)[r]{\strut{} 2}}%
      \put(528,2599){\makebox(0,0)[r]{\strut{} 2.5}}%
      \put(660,484){\makebox(0,0){\strut{} 30}}%
      \put(1208,484){\makebox(0,0){\strut{} 40}}%
      \put(1756,484){\makebox(0,0){\strut{} 50}}%
      \put(2305,484){\makebox(0,0){\strut{} 60}}%
      \put(2853,484){\makebox(0,0){\strut{} 70}}%
      \put(3401,484){\makebox(0,0){\strut{} 80}}%
      \put(3949,484){\makebox(0,0){\strut{} 90}}%
      \put(22,1721){\rotatebox{-270}{\makebox(0,0){\strut{}Core Position [b]}}}%
      \put(2304,154){\makebox(0,0){\strut{}$(10\bar{1})$ plane index}}%
    \put(0,0){\includegraphics{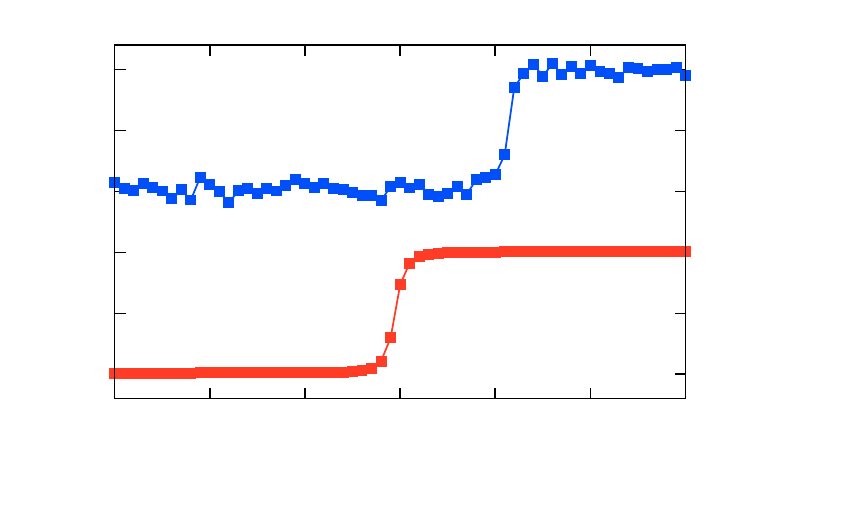}}%
  \end{picture}%
\caption{The relaxed kink profile at zero (Red) and finite temperature (Blue). The centre of mass for the kink has propagated but remains sharp.\label{kinkprof}}
\end{figure}
To do this, we obtain first the kink equation of motion, assuming that the kink position $X(t)$ is slowly varying in time in comparison to the fluctuations of the $\{u_n\}$, and that the kink shape suffers only small perturbations due to these fluctuations throughout its motion. It will be seen that these approximations still give excellent agreement with the results from stochastic integration, and allow us to equate the field center of mass to the kink position $X(t)$. To derive the kink equation of motion we evaluate the total velocity projected along the dislocation line.  We consider segments connecting neighboring nodes of the dislocation line and take the projection of the nodal velocity along the segment normal. We then take the projection of this normal along the line direction $\bf \hat x$ to obtain the contribution to the net projection from that segment. A very similar calculation is used when calculating the driving force on a ship's sail\cite{sail} and is formally equivalent to the derivation of the field momentum of (\ref{DV}) in the continuum limit\cite{Kosevich2006}. It can be shown that this gives a kink velocity
\begin{equation}
\dot{X}(t) = \sum_n{d u_n(t) \over dt} \left(u_{n+1}(t)-u_{n}(t)\over a\right).
\end{equation}
Using (\ref{lEoM}), (\ref{KMP}) and (\ref{FKkink}) we obtain the kink equation of motion
\begin{equation}
\gamma_{kink}\dot{X}(t)={\pi\over a}E_{mig}\sin{2\pi X(t)\over a} + \eta(t),\label{dKEL}
\end{equation}
where the kink dissipation parameter $\gamma_{kink}$ is given by
\begin{equation}
\gamma_{kink} = {L_P\over4\pi w_0}\gamma_{line},\label{gamma_kink}
\end{equation}
$\eta(t)$ is a one dimensional Gaussian random variable with an average and standard deviation
\begin{equation}
\langle \eta(t)\rangle = 0,\quad\langle\eta(t)\eta(t')\rangle = 2\gamma_{kink}T\delta(t'-t),
\end{equation}
and $E_{mig}$ is given in equation (\ref{kink_energy}). We note that equation (\ref{gamma_kink}) shows the dissipation parameter for the dislocation line, $\gamma_{line}$, to be directly proportional to the dissipation parameter for the kink, $\gamma_{kink}$, and thus $\gamma_{line}$ is also temperature independent. Equation (\ref{dKEL}) describes a point particle undergoing one dimensional stochastic motion in a sinusoidal potential. As discussed in section \ref{sec:MD}, under an ensemble average the mean squared displacement exhibits diffusive behavior with a diffusion constant (\ref{BessD}), where now $E_{mig}$ and $\gamma_{kink}$ are given explicitly in terms of the free model parameters $V_P,\gamma_{line},\kappa$ and the crystallographically determined $L_P,a$. By inverting the relations (\ref{FKkink}), (\ref{kink_energy}) and (\ref{gamma_kink}) we may determine the model parameters for the dislocations considered here. With these parameters, the system (\ref{DV}) was first relaxed by a conjugate gradient algorithm, with the boundary conditions (\ref{KBC}), to determine the formation energy and kink width, then the equations of motion (\ref{lEoM}) were stochastically integrated. The kink formed remains well defined at finite temperature, as shown in Figure \ref{kinkprof}. It can be seen that the data is identical in form to that produced from atomistic simulation, shown in Figure \ref{110_kink}, with the kink trajectories extracted and analyzed in an identical manner. Table \ref{Kink_Table} shows the results from these simulations as compared to the results from atomistic simulation, displaying excellent agreement over a wide range of temperature.\\

It is emphasized that the discreteness of the model is essential to produce a kink migration barrier. We also note that the discreteness, which is determined by the underlying crystallography through $a$ and $L_P$, influences the kink formation energy as well as the migration barrier, as can be seen in equation (\ref{kink_energy}). As we simulate a line of only $\sim$$500$ nodes on a coarse time step of $10$ ps, as opposed to the entire atomistic system of $700,000$ atoms on a very fine time step of $1$ fs, we may generate data sets equivalent to those produced from atomistic simulation at around $\sim$$10^{-7}$ of the computational cost. Therefore, despite the atomistic resolution along the line, the model affords enormous computational savings as compared with a full MD simulation. This significant efficiency gain allows us to simulate dislocation motion at experimental strain rates, while retaining atomistic resolution and a statistically rigorous temperature.

\begin{table}[!h]
\begin{tabular*}{0.48\textwidth}{@{\extracolsep{\fill}}ccccc}\toprule
Dislocation & Simulation & $E_{kink}$ & $E_{mig}$ & $\gamma_{kink}$ \\
System & Method & [eV] & [eV] & [m$_u$ps$^{-1}$]\\
\colrule
${1/2}[111](1\bar{2}1)$ Edge& MD & 0.150 & 0.074 & 1.79 \\
	& FKL & 0.148 & 0.072 & 1.74 \\
$[100](010)$ Edge& MD & 0.510 & 0.222 & 2.61 \\
& FKL & 0.505 & 0.218 & 2.58 \\
${1/2}[111](1\bar{1}0)$ Screw& MD & 0.367 & - & 1.83 \\
{(Average)}& FKL & 0.367 & - & 1.82 \\
\botrule
\end{tabular*}
\caption{Formation energies, migration energies and dissipation parameters obtained from  MD and FKL simulations for kinks on the dislocations investigated here. The values were obtained by identical processing for each simulation technique. The MD data for kinks on screw dislocations is the average between left and right kinks as detailed in the text. Very good agreement between the MD and FKL parameters is seen.\label{Kink_Table}}
\end{table}\subsection{Dislocation motion under vanishing applied stress}
The parameters obtained from the kink diffusion simulations are now used to investigate the motion of straight dislocations at experimental stress levels. This important regime is not accessible to atomistic simulation for dislocations which have a large kink formation energy. Therefore, this is an ideal application of the FKL model. For a discrete dislocation segment of $N$ nodes, we supplement the equations of motion (\ref{lEoM}) with a force per node $f$ to induce drift of the dislocation line,
\begin{equation}
\gamma_{line}{d u_n(t) \over dt}=-{\partial\over\partial u_n}V(\{u_m\}) + f + \eta_n(t), \label{dlEoM}
\end{equation}
with periodic boundary conditions
\begin{equation}
u_n(t) = u_{n+N}(t).
\end{equation}
We then extract the position of the center of mass ${\bar u} = \sum_n {u}_{n}/N$ at each timestep, obtaining the ensemble average center of mass velocity $\langle{\bar v}\rangle$ in an identical manner to that shown in equation (\ref{MDMSD}),
\begin{equation}
\langle{\bar v}\rangle = \sum_{n=0}^{N-\tau/\Delta{t}}{(\bar{u}_{n\Delta{t}+\tau}-{\bar u}_{n\Delta{t}})\over{\tau(N-\tau/\Delta{t})}}.\label{drift_FKL}
\end{equation}
To obtain the relationship between the force $f$ and an applied stress, we recall the Peach-Koehler formula\cite{Hirth} for the force per unit length ${\bf f}_{PK}$ on a dislocation of Burgers vector $\bf b$ and line direction $\bf t$ under an applied stress $\bf \sigma$,
\begin{equation}
{\bf f}_{PK} = ({\bf \sigma}\cdot{\bf b})\wedge{\bf \hat t}.\label{PK}
\end{equation}
The nodal force $f$ is then the projection of (\ref{PK}) along the displacement direction of the $\{u_n\}$, ${\bf \hat u}$, multiplied by the segment separation $a$. We apply a shear stress across the slip plane of magnitude $|\sigma|$ in the direction of the dislocation Burgers vector $\bf b$, resulting in a force per node of
\begin{equation}
f = a\space{\bf f}_{PK}\cdot{\bf \hat u} = a|{\bf b}||\sigma|.
\end{equation}

{{We apply experimental stresses of 40 MPa, which corresponds to a very small force per node of $\sim$$10^{-3}$eV/$L_P$.  To demonstrate phenomena this discrete model can treat, we investigate the effect of segment length on dislocation velocity. Figures \ref{M_E_S},\ref{M_E_D} show typical center of mass trajectories of ${1/2}[111](1\bar{1}0)$ screw and $[100](010)$ edge dislocation segments. Extracting the center of mass velocity through equation (\ref{drift_FKL}) over a wide range of segment lengths gave a length independent velocity for edge segments, whereas the velocity increased linearly with segment length as $0.013(4)s^{-1}$ at 300K for screw dislocations. This is in good agreement with the linear relationship of velocity with length of around $0.01s^{-1}$ at 300K for screw dislocation segments that has recently been observed experimentally\cite{Caillard2}. We note that this gradient depends exponentially on temperature due to the activated nature of the dislocation migration.\\
We can understand these differences in terms of the discrete structure of the FKL dislocation lines. We have seen that the wide kinks on screw dislocations have a negligible migration barrier; as a result, once a double kink is formed it will move quickly under the applied stress until it meets another kink. For long segments we therefore expect the dislocation velocity to scale linearly with segment length due the increased number of possible locations for kink nucleation, as observed. In contrast, as the narrow kinks on edge dislocations have a large kink migration barrier, comparable to the kink formation energy. As a result we expect the mobility always to be independent of the segment length as kink migration and kink nucleation occur on similar timescales. This behavior, which is found only in a discrete line model, is expected to have consequences in many aspects of dislocation behavior, for example the difference in the effect of impurities on the mobilities of edge and screw dislocations\cite{Hirth}.}}
\begin{figure}
  \setlength{\unitlength}{0.0500bp}%
  \begin{picture}(4874.00,3004.00)%
      \put(528,704){\makebox(0,0)[r]{\strut{} 0}}%
      \put(528,1017){\makebox(0,0)[r]{\strut{} 5}}%
      \put(528,1330){\makebox(0,0)[r]{\strut{} 10}}%
      \put(528,1643){\makebox(0,0)[r]{\strut{} 15}}%
      \put(528,1956){\makebox(0,0)[r]{\strut{} 20}}%
      \put(528,2269){\makebox(0,0)[r]{\strut{} 25}}%
      \put(528,2582){\makebox(0,0)[r]{\strut{} 30}}%
      \put(660,484){\makebox(0,0){\strut{} 0}}%
      \put(1548,484){\makebox(0,0){\strut{} .5}}%
      \put(2437,484){\makebox(0,0){\strut{} 1}}%
      \put(3325,484){\makebox(0,0){\strut{} 1.5}}%
      \put(4213,484){\makebox(0,0){\strut{} 2}}%
      \put(90,1721){\rotatebox{-270}{\makebox(0,0){\strut{}Center of Mass $\bar u$ [${\rm\AA}$]}}}%
      \put(2436,154){\makebox(0,0){\strut{}Time [s]}}%
      \put(1415,2087){\makebox(0,0)[l]{\strut{}L $\simeq$ 200nm, $\dot{\bar{u}}\sim$16${\rm\AA}s^{-1}$}}%
      \put(2081,917){\makebox(0,0)[l]{\strut{}L $\simeq$ 100nm, $\dot{\bar{u}}\sim$8${\rm\AA}s^{-1}$}}%
      \put(0,0){\includegraphics{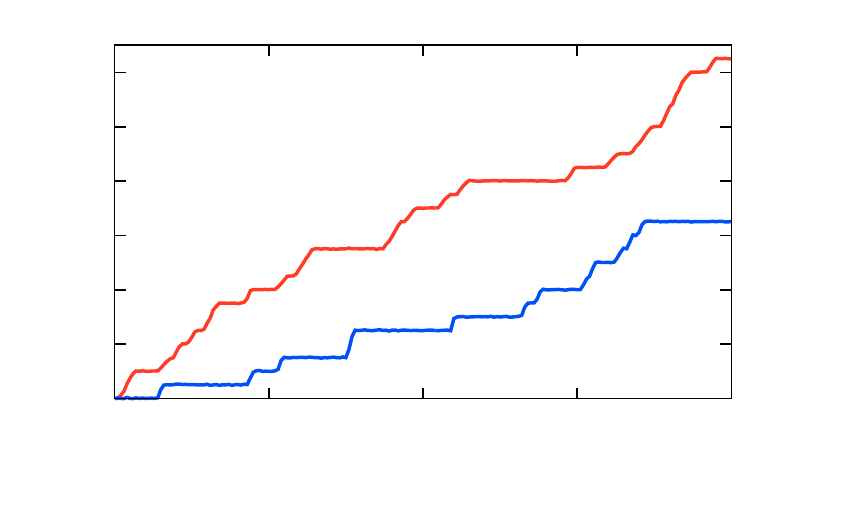}}%
  \end{picture}%

\caption{{Center of mass positions for two ${1/2}[111](1\bar{1}0)$ screw dislocation segments, under an applied stress of 40 MPa, at a temperature of 300K. The highly stepped motion reflects directly the nucleation of kinks, which then quickly propagate along the entire line due to the negligible kink migration barrier. As a result longer lines, which have a greater number of nucleation sites, have a velocity which increases linearly with segment length.}\label{M_E_S}}
\end{figure}
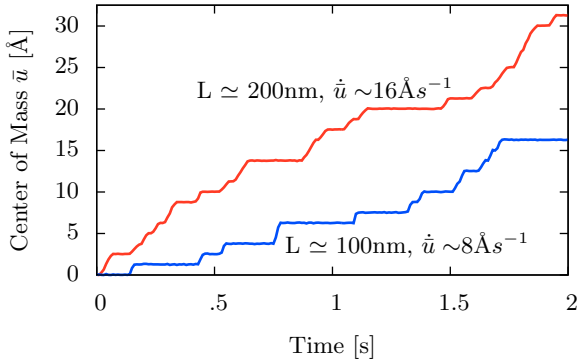
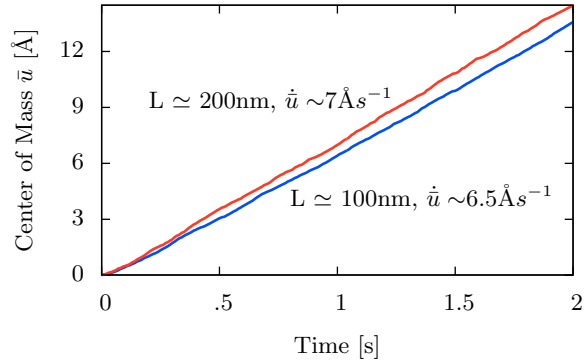
\begin{figure}
  \setlength{\unitlength}{0.0500bp}%
  \begin{picture}(4874.00,3004.00)%
      \put(528,702){\makebox(0,0)[r]{\strut{} 0}}%
      \put(528,1124){\makebox(0,0)[r]{\strut{} 3}}%
      \put(528,1546){\makebox(0,0)[r]{\strut{} 6}}%
      \put(528,1968){\makebox(0,0)[r]{\strut{} 9}}%
      \put(528,2390){\makebox(0,0)[r]{\strut{} 12}}%
      \put(660,484){\makebox(0,0){\strut{} 0}}%
      \put(1548,484){\makebox(0,0){\strut{} .5}}%
      \put(2437,484){\makebox(0,0){\strut{} 1}}%
      \put(3325,484){\makebox(0,0){\strut{} 1.5}}%
      \put(4213,484){\makebox(0,0){\strut{} 2}}%
      \put(90,1721){\rotatebox{-270}{\makebox(0,0){\strut{}Center of Mass $\bar u$ [${\rm\AA}$]}}}%
      \put(2436,154){\makebox(0,0){\strut{}Time [s]}}%
      \put(1015,2007){\makebox(0,0)[l]{\strut{}L $\simeq$ 200nm, $\dot{\bar{u}}\sim$7${\rm\AA} s^{-1}$ }}%
      \put(2081,1274){\makebox(0,0)[l]{\strut{}L $\simeq$ 100nm, $\dot{\bar{u}}\sim$6.5${\rm\AA} s^{-1}$}}%
    \put(0,0){\includegraphics{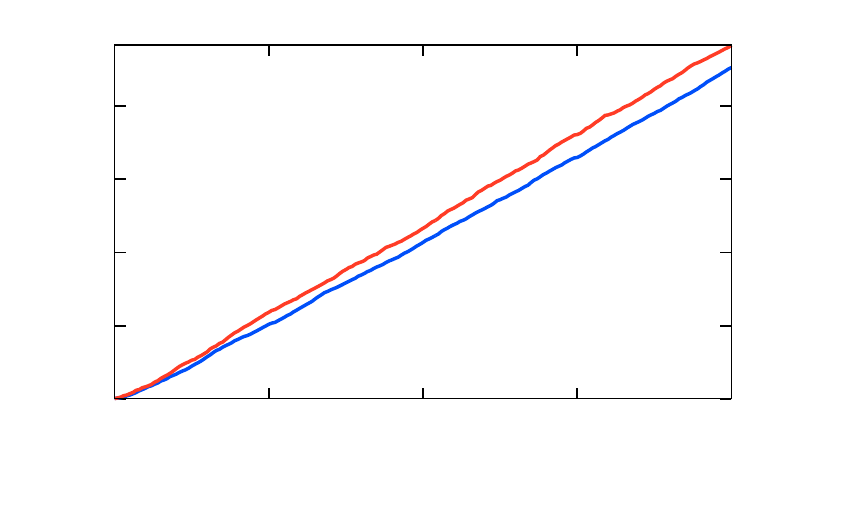}}%
  \end{picture}%
\caption{{Center of mass positions for two $[100](010)$ edge dislocation segments, under an applied stress of 40 MPa, at a temperature of 600K. As the kink migration barrier is comaparable to the kink formation energy double kink nucleation occurs on the same timescale as kink migration and thus the line propagates gradually. As a result the segment velocity is almost independent of the segment length, as distinct from screw dislocation segments, a feature not captured by a continuum line model.}\label{M_E_D}}
\end{figure}\section{Conclusions}\label{sec:disc}
In this paper we have presented results from large scale MD simulations of edge and screw dislocation lines in bcc Fe. Boundary conditions were exploited to produce isolated kinks, which were seen to remain isolated on the timescale of the MD simulations provided the thermal energy was significantly less than the double kink formation energy. The dislocation lines were coarse grained, while retaining atomistic resolution, by assigning a dislocation core position in each atomic plane normal to the line direction. Kinks were clearly identifiable, with statistical analysis allowing extraction of a kink diffusion constant.\\

Kinks on edge dislocations were seen to be narrow, exhibiting thermally activated stochastic motion which was described well by an Arrhenius law. In contrast, kinks on screw dislocations were broad with a diffusivity that varied linearly with temperature, implying a vanishingly small migration barrier. The difference in the formation energies of left and right kinks on screw dislocations was seen to be attributable to the structure of the kinks outside their cores. The kink interaction energy predicted by elasticity was observed to be much smaller than the kink core energy.\\

An analytic result for one dimensional stochastic motion in a periodic potential, valid for all temperatures and barrier heights, was seen to rationalize the wide range of diffusive behavior, leading to the conclusion that the frictional force on a kink, and hence the host dislocation, is temperature independent. This result from direct atomistic simulation is in agreement with other studies on dislocations with a large lattice resistance, but directly opposes the textbook theory which states that the frictional force should be proportional to temperature. This significant disagreement is an important topic for future investigation.\\

The discrete one-dimensional FKL model was seen to be able to reproduce the observed motion of kinks on edge and screw dislocations over a wide range of temperatures and dislocation geometries, with all length scales fixed by the crystallography. The discrepancy between kinks on edge and screw dislocations was explained through an analytic relationship between the kink width and migration barrier; it was noted that the migration barrier depends sensitively on the discrete structure, with the atomistic resolution of the model being essential to reproduce the detailed behavior of thermally activated dislocation glide. This discrete structure, which is absent in conventional dislocation dynamics codes, is expected to be significant in thermally activated dislocation phenomena such as impurity interaction and cross slip.\\

The application of the FKL model to the motion of initially straight dislocation segments under experimental applied stresses found a noticeable length dependence for screw dislocation segments, due to the negligible kink migration barrier, whereas the large migration barrier for kinks on edge dislocations suppressed a length dependence. {{The consequences of this highly anisotropic mobility of discrete dislocation lines for microstructural evolution will be investigated in future work.}}
\section{Acknowledgements}
{{We are grateful to a referee for helpful comments on an earlier draft of this paper.}} TDS was supported through a studentship in the Centre for Doctoral Training on Theory and Simulation of Materials at Imperial College London funded by EPSRC under grant number EP/G036888/1. This work, partially supported by the European Communities under the contract of Association between EURATOM and CCFE, was carried out within the framework of the European Fusion Development Agreement. The views and opinions expressed herein do not necessarily reflect those of the European Commission. This work was also part-funded by the RCUK Energy Programme under grant EP/I501045.


\begin{thebibliography}{52}
\expandafter\ifx\csname natexlab\endcsname\relax\def\natexlab#1{#1}\fi
\expandafter\ifx\csname bibnamefont\endcsname\relax
  \def\bibnamefont#1{#1}\fi
\expandafter\ifx\csname bibfnamefont\endcsname\relax
  \def\bibfnamefont#1{#1}\fi
\expandafter\ifx\csname citenamefont\endcsname\relax
  \def\citenamefont#1{#1}\fi
\expandafter\ifx\csname url\endcsname\relax
  \def\url#1{\texttt{#1}}\fi
\expandafter\ifx\csname urlprefix\endcsname\relax\def\urlprefix{URL }\fi
\providecommand{\bibinfo}[2]{#2}
\providecommand{\eprint}[2][]{\url{#2}}

\bibitem[{\citenamefont{Argon}(2008)}]{Argon}
\bibinfo{author}{\bibfnamefont{A.~S.} \bibnamefont{Argon}},
  \emph{\bibinfo{title}{Strengthening Mechanisms in Crystal Plasticity}},
  Oxford Series on Materials Modelling (\bibinfo{publisher}{Oxford Univ.
  Press}, \bibinfo{address}{Oxford}, \bibinfo{year}{2008}).

\bibitem[{\citenamefont{{Queyreau} et~al.}(2011)\citenamefont{{Queyreau},
  {Marian}, {Gilbert}, and {Wirth}}}]{Gilbert2011}
\bibinfo{author}{\bibfnamefont{S.}~\bibnamefont{{Queyreau}}},
  \bibinfo{author}{\bibfnamefont{J.}~\bibnamefont{{Marian}}},
  \bibinfo{author}{\bibfnamefont{M.~R.} \bibnamefont{{Gilbert}}},
  \bibnamefont{and} \bibinfo{author}{\bibfnamefont{B.~D.}
  \bibnamefont{{Wirth}}}, \bibinfo{journal}{\prb}
  {\bibinfo{volume}{84}}, \bibinfo{eid}{064106} (\bibinfo{year}{2011}).

\bibitem[{\citenamefont{Monnet and Terentyev}(2009)}]{Terentyev2009}
\bibinfo{author}{\bibfnamefont{G.}~\bibnamefont{Monnet}} \bibnamefont{and}
  \bibinfo{author}{\bibfnamefont{D.}~\bibnamefont{Terentyev}},
  \bibinfo{journal}{Acta Materialia} {\bibinfo{volume}{57}},
  \bibinfo{pages}{1416 } (\bibinfo{year}{2009}).

\bibitem[{\citenamefont{Rodney and Proville}(2009)}]{Rodney2009}
\bibinfo{author}{\bibfnamefont{D.}~\bibnamefont{Rodney}} \bibnamefont{and}
  \bibinfo{author}{\bibfnamefont{L.}~\bibnamefont{Proville}},
  \bibinfo{journal}{Phys. Rev. B} {\bibinfo{volume}{79}},
  \bibinfo{pages}{094108} (\bibinfo{year}{2009}).

\bibitem[{\citenamefont{Leibfried}(1950)}]{Leibfried}
\bibinfo{author}{\bibfnamefont{G.}~\bibnamefont{Leibfried}},
  \bibinfo{journal}{Z. Phys} {\bibinfo{volume}{127}},
  \bibinfo{pages}{344} (\bibinfo{year}{1950}).

\bibitem[{\citenamefont{Gilbert et~al.}(2011)\citenamefont{Gilbert, Queyreau,
  and Marian}}]{Gilbert}
\bibinfo{author}{\bibfnamefont{M.~R.} \bibnamefont{Gilbert}},
  \bibinfo{author}{\bibfnamefont{S.}~\bibnamefont{Queyreau}}, \bibnamefont{and}
  \bibinfo{author}{\bibfnamefont{J.}~\bibnamefont{Marian}},
  \bibinfo{journal}{Phys. Rev. B} {\bibinfo{volume}{84}},
  \bibinfo{pages}{174103} (\bibinfo{year}{2011}).

\bibitem[{\citenamefont{Dudarev}(2002)\citenamefont{Dudarev}}]{Dudarev2002}
\bibinfo{author}{\bibfnamefont{S.~L.} \bibnamefont{Dudarev}},
  \bibinfo{journal}{Phys. Rev. B} {\bibinfo{volume}{65}},
  \bibinfo{pages}{224105} (\bibinfo{year}{2002}).

\bibitem[{\citenamefont{Dudarev}(2008)}]{Dudarev2008b}
\bibinfo{author}{\bibfnamefont{S.~L.} \bibnamefont{Dudarev}},
  \bibinfo{journal}{Comptes Rendus Physique} {\bibinfo{volume}{9}},
  \bibinfo{pages}{409 } (\bibinfo{year}{2008}).

\bibitem[{\citenamefont{Marian et~al.}(2002)\citenamefont{Marian, Wirth, Caro,
  Sadigh, Odette, Perlado, and Diaz de~la Rubia}}]{Marian2002}
\bibinfo{author}{\bibfnamefont{J.}~\bibnamefont{Marian}},
  \bibinfo{author}{\bibfnamefont{B.~D.} \bibnamefont{Wirth}},
  \bibinfo{author}{\bibfnamefont{A.}~\bibnamefont{Caro}},
  \bibinfo{author}{\bibfnamefont{B.}~\bibnamefont{Sadigh}},
  \bibinfo{author}{\bibfnamefont{G.~R.} \bibnamefont{Odette}},
  \bibinfo{author}{\bibfnamefont{J.~M.} \bibnamefont{Perlado}},
  \bibnamefont{and} \bibinfo{author}{\bibfnamefont{T.}~\bibnamefont{Diaz de~la
  Rubia}}, \bibinfo{journal}{Phys. Rev. B} {\bibinfo{volume}{65}},
  \bibinfo{pages}{144102} (\bibinfo{year}{2002}).

\bibitem[{\citenamefont{Braun and Kivshar}(1998)}]{Braun1998}
\bibinfo{author}{\bibfnamefont{O.~M.} \bibnamefont{Braun}} \bibnamefont{and}
  \bibinfo{author}{\bibfnamefont{Y.~S.} \bibnamefont{Kivshar}},
  \bibinfo{journal}{Physics Reports} {\bibinfo{volume}{306}},
  \bibinfo{pages}{1 } (\bibinfo{year}{1998}), ISSN \bibinfo{issn}{0370-1573}.

\bibitem[{\citenamefont{Seeger}(1999)}]{Seeger}
\bibinfo{author}{\bibfnamefont{A.}~\bibnamefont{Seeger}}, in
  \emph{\bibinfo{booktitle}{Physical Acoustics: Principles and Methods}},
  edited by \bibinfo{editor}{\bibfnamefont{W.}~\bibnamefont{Mason}}
  \bibnamefont{and} \bibinfo{editor}{\bibfnamefont{R.}~\bibnamefont{Thurston}}
  (\bibinfo{publisher}{Academic Press}, \bibinfo{year}{1999}),
  vol.~\bibinfo{volume}{3}, chap.~\bibinfo{chapter}{7}, ISBN
  \bibinfo{isbn}{9780124779457}.

\bibitem[{\citenamefont{Hirth and Lothe}(1991)}]{Hirth}
\bibinfo{author}{\bibfnamefont{J.~P.} \bibnamefont{Hirth}} \bibnamefont{and}
  \bibinfo{author}{\bibfnamefont{J.}~\bibnamefont{Lothe}},
  \emph{\bibinfo{title}{Theory Of Dislocations}} (\bibinfo{publisher}{Krieger,
  Malabar, FL}, \bibinfo{year}{1991}).

\bibitem[{\citenamefont{Bulatov and Cai}(2003)}]{BulatovBook}
\bibinfo{author}{\bibfnamefont{V.}~\bibnamefont{Bulatov}} \bibnamefont{and}
  \bibinfo{author}{\bibfnamefont{W.}~\bibnamefont{Cai}},
  \emph{\bibinfo{title}{Computer Simulations Of Dislocations}}
  (\bibinfo{publisher}{Oxford University Press}, \bibinfo{year}{2003}).

\bibitem[{\citenamefont{Derlet et~al.}(2011)\citenamefont{Derlet, Gilbert, and
  Dudarev}}]{Dudarev2011}
\bibinfo{author}{\bibfnamefont{P.~M.} \bibnamefont{Derlet}},
  \bibinfo{author}{\bibfnamefont{M.~R.} \bibnamefont{Gilbert}},
  \bibnamefont{and} \bibinfo{author}{\bibfnamefont{S.~L.}
  \bibnamefont{Dudarev}}, \bibinfo{journal}{Phys. Rev. B}
  {\bibinfo{volume}{84}}, \bibinfo{pages}{134109}
  (\bibinfo{year}{2011}).

\bibitem[{\citenamefont{Plimpton}(1995)}]{LAMMPS}
\bibinfo{author}{\bibfnamefont{S.}~\bibnamefont{Plimpton}},
  \bibinfo{journal}{Journal Computational Physics}
  {\bibinfo{volume}{117}}, \bibinfo{pages}{1} (\bibinfo{year}{1995}).

\bibitem[{\citenamefont{Gordon et~al.}(2011)\citenamefont{Gordon, Neeraj, and
  Mendelev}}]{NewMendelev}
\bibinfo{author}{\bibfnamefont{P.}~\bibnamefont{Gordon}},
  \bibinfo{author}{\bibfnamefont{T.}~\bibnamefont{Neeraj}}, \bibnamefont{and}
  \bibinfo{author}{\bibfnamefont{M.}~\bibnamefont{Mendelev}},
  \bibinfo{journal}{Philosophical Magazine} {\bibinfo{volume}{91}},
  \bibinfo{pages}{3931} (\bibinfo{year}{2011}).

\bibitem[{\citenamefont{Clouet et~al.}(2009)\citenamefont{Clouet, Ventelon, and
  Willaime}}]{DFTScrew}
\bibinfo{author}{\bibfnamefont{E.}~\bibnamefont{Clouet}},
  \bibinfo{author}{\bibfnamefont{L.}~\bibnamefont{Ventelon}}, \bibnamefont{and}
  \bibinfo{author}{\bibfnamefont{F.}~\bibnamefont{Willaime}},
  \bibinfo{journal}{Phys. Rev. Lett.} {\bibinfo{volume}{102}},
  \bibinfo{pages}{055502} (\bibinfo{year}{2009}).

\bibitem[{\citenamefont{Bulatov}(1997)}]{Bulatov97}
\bibinfo{author}{\bibfnamefont{V.}~\bibnamefont{Bulatov}},
  \bibinfo{journal}{Physical Review Letters} {\bibinfo{volume}{79}},
  \bibinfo{pages}{5042} (\bibinfo{year}{1997}).

\bibitem[{\citenamefont{Wang et~al.}(2003)\citenamefont{Wang, Strachan, Cagin,
  and Goddard}}]{Wang2003}
\bibinfo{author}{\bibfnamefont{G.}~\bibnamefont{Wang}},
  \bibinfo{author}{\bibfnamefont{A.}~\bibnamefont{Strachan}},
  \bibinfo{author}{\bibfnamefont{T.}~\bibnamefont{Cagin}}, \bibnamefont{and}
  \bibinfo{author}{\bibfnamefont{W.~A.} \bibnamefont{Goddard}},
  \bibinfo{journal}{Phys. Rev. B} {\bibinfo{volume}{68}},
  \bibinfo{pages}{224101} (\bibinfo{year}{2003}).

\bibitem[{\citenamefont{Ventelon et~al.}(2009)\citenamefont{Ventelon, Willaime,
  and Leyronnas}}]{Ventelon2009}
\bibinfo{author}{\bibfnamefont{L.}~\bibnamefont{Ventelon}},
  \bibinfo{author}{\bibfnamefont{F.}~\bibnamefont{Willaime}}, \bibnamefont{and}
  \bibinfo{author}{\bibfnamefont{P.}~\bibnamefont{Leyronnas}},
  \bibinfo{journal}{Journal of Nuclear Materials}
  {\bibinfo{volume}{386-388}}, \bibinfo{pages}{26}
  (\bibinfo{year}{2009}).

\bibitem[{\citenamefont{Clouet et~al.}(2008)\citenamefont{Clouet, Garruchet,
  Nguyen, Perez, and Becquart}}]{Clouet2008}
\bibinfo{author}{\bibfnamefont{E.}~\bibnamefont{Clouet}},
  \bibinfo{author}{\bibfnamefont{S.}~\bibnamefont{Garruchet}},
  \bibinfo{author}{\bibfnamefont{H.}~\bibnamefont{Nguyen}},
  \bibinfo{author}{\bibfnamefont{M.}~\bibnamefont{Perez}}, \bibnamefont{and}
  \bibinfo{author}{\bibfnamefont{C.~S.} \bibnamefont{Becquart}},
  \bibinfo{journal}{Acta Materialia} {\bibinfo{volume}{56}},
  \bibinfo{pages}{3450 } (\bibinfo{year}{2008}), ISSN
  \bibinfo{issn}{1359-6454}.

\bibitem[{\citenamefont{Li-Qun et~al.}(2008)\citenamefont{Li-Qun, Chong-Yu, and
  Tao}}]{LiQun}
\bibinfo{author}{\bibfnamefont{C.}~\bibnamefont{Li-Qun}},
  \bibinfo{author}{\bibfnamefont{W.}~\bibnamefont{Chong-Yu}}, \bibnamefont{and}
  \bibinfo{author}{\bibfnamefont{Y.}~\bibnamefont{Tao}},
  \bibinfo{journal}{Chinese Physics B} {\bibinfo{volume}{17}},
  \bibinfo{pages}{662} (\bibinfo{year}{2008}).

\bibitem[{\citenamefont{Chang et~al.}(2001)\citenamefont{Chang, Cai, Bulatov,
  and Yip}}]{Chang}
\bibinfo{author}{\bibfnamefont{J.}~\bibnamefont{Chang}},
  \bibinfo{author}{\bibfnamefont{W.}~\bibnamefont{Cai}},
  \bibinfo{author}{\bibfnamefont{V.}~\bibnamefont{Bulatov}}, \bibnamefont{and}
  \bibinfo{author}{\bibfnamefont{S.}~\bibnamefont{Yip}},
  \bibinfo{journal}{Materials Science and Engineering A}
  (\bibinfo{year}{2001}).

\bibitem[{\citenamefont{Dudarev et~al.}(2010)\citenamefont{Dudarev, Gilbert,
  Arakawa, Mori, Yao, Jenkins, and Derlet}}]{Dudarev2010}
\bibinfo{author}{\bibfnamefont{S.~L.} \bibnamefont{Dudarev}},
  \bibinfo{author}{\bibfnamefont{M.~R.} \bibnamefont{Gilbert}},
  \bibinfo{author}{\bibfnamefont{K.}~\bibnamefont{Arakawa}},
  \bibinfo{author}{\bibfnamefont{H.}~\bibnamefont{Mori}},
  \bibinfo{author}{\bibfnamefont{Z.}~\bibnamefont{Yao}},
  \bibinfo{author}{\bibfnamefont{M.~L.} \bibnamefont{Jenkins}},
  \bibnamefont{and} \bibinfo{author}{\bibfnamefont{P.~M.}
  \bibnamefont{Derlet}}, \bibinfo{journal}{Phys. Rev. B}
  {\bibinfo{volume}{81}}, \bibinfo{pages}{224107}
  (\bibinfo{year}{2010}).

\bibitem[{\citenamefont{Reichl}(2009)}]{Reichl2009}
\bibinfo{author}{\bibfnamefont{L.}~\bibnamefont{Reichl}},
  \emph{\bibinfo{title}{A Modern Course in Statistical Physics}}, Physics
  Textbook (\bibinfo{publisher}{Wiley-VCH}, \bibinfo{year}{2009}), ISBN
  \bibinfo{isbn}{9783527407828}.

\bibitem[{\citenamefont{{Lifson} and {Jackson}}(1962)}]{Lifson}
\bibinfo{author}{\bibfnamefont{S.}~\bibnamefont{{Lifson}}} \bibnamefont{and}
  \bibinfo{author}{\bibfnamefont{J.~L.} \bibnamefont{{Jackson}}},
  \bibinfo{journal}{\jcp} {\bibinfo{volume}{36}}, \bibinfo{pages}{2410}
  (\bibinfo{year}{1962}).

\bibitem[{\citenamefont{Coffey et~al.}(2004)\citenamefont{Coffey, Kalmykov, and
  Waldron}}]{coffey2004}
\bibinfo{author}{\bibfnamefont{W.}~\bibnamefont{Coffey}},
  \bibinfo{author}{\bibfnamefont{Y.}~\bibnamefont{Kalmykov}}, \bibnamefont{and}
  \bibinfo{author}{\bibfnamefont{J.}~\bibnamefont{Waldron}},
  \emph{\bibinfo{title}{The Langevin Equation: With Applications to Stochastic
  Problems in Physics, Chemistry, and Electrical Engineering}}, World
  Scientific Series in Contemporary Chemical Physics (\bibinfo{publisher}{World
  Scientific}, \bibinfo{year}{2004}), ISBN \bibinfo{isbn}{9789812384621}.

\bibitem[{\citenamefont{{H{\"a}nggi} et~al.}(1990)\citenamefont{{H{\"a}nggi},
  {Talkner}, and {Borkovec}}}]{Hanggi}
\bibinfo{author}{\bibfnamefont{P.}~\bibnamefont{{H{\"a}nggi}}},
  \bibinfo{author}{\bibfnamefont{P.}~\bibnamefont{{Talkner}}},
  \bibnamefont{and}
  \bibinfo{author}{\bibfnamefont{M.}~\bibnamefont{{Borkovec}}},
  \bibinfo{journal}{Reviews of Modern Physics} {\bibinfo{volume}{62}},
  \bibinfo{pages}{251} (\bibinfo{year}{1990}).

\bibitem[{\citenamefont{Einstein and F{\"u}rth}(1956)}]{einstein1956}
\bibinfo{author}{\bibfnamefont{A.}~\bibnamefont{Einstein}} \bibnamefont{and}
  \bibinfo{author}{\bibfnamefont{R.}~\bibnamefont{F{\"u}rth}},
  \emph{\bibinfo{title}{Investigations on the Theory of the Brownian
  Movement}}, Dover Books on Physics (\bibinfo{publisher}{Dover Publications},
  \bibinfo{year}{1956}), ISBN \bibinfo{isbn}{9780486603049}.

\bibitem[{\citenamefont{Abramowitz and Stegun}(1965)}]{Abram}
\bibinfo{author}{\bibfnamefont{M.}~\bibnamefont{Abramowitz}} \bibnamefont{and}
  \bibinfo{author}{\bibfnamefont{I.}~\bibnamefont{Stegun}},
  \emph{\bibinfo{title}{Handbook of Mathematical Functions: With Formulas,
  Graphs, and Mathematical Tables}}, Applied mathematics series
  (\bibinfo{publisher}{Dover Publications}, \bibinfo{year}{1965}), ISBN
  \bibinfo{isbn}{9780486612720}.

\bibitem[{\citenamefont{Rong et~al.}(2005)\citenamefont{Rong, Osetsky, and
  Bacon}}]{Rong}
\bibinfo{author}{\bibfnamefont{Z.}~\bibnamefont{Rong}},
  \bibinfo{author}{\bibfnamefont{Y.~N.} \bibnamefont{Osetsky}},
  \bibnamefont{and} \bibinfo{author}{\bibfnamefont{D.~J.} \bibnamefont{Bacon}},
  \bibinfo{journal}{Philosophical Magazine} {\bibinfo{volume}{85}},
  \bibinfo{pages}{1473} (\bibinfo{year}{2005}).

\bibitem[{\citenamefont{Bacon et~al.}(2009)\citenamefont{Bacon, Osetsky, and
  Rodney}}]{RodneyBacon}
\bibinfo{author}{\bibfnamefont{D.}~\bibnamefont{Bacon}},
  \bibinfo{author}{\bibfnamefont{Y.}~\bibnamefont{Osetsky}}, \bibnamefont{and}
  \bibinfo{author}{\bibfnamefont{D.}~\bibnamefont{Rodney}}, in
  \emph{\bibinfo{booktitle}{Dislocations Obstacle interactions at the atomic
  level}}, edited by \bibinfo{editor}{\bibfnamefont{J.}~\bibnamefont{Hirth}}
  \bibnamefont{and} \bibinfo{editor}{\bibfnamefont{L.}~\bibnamefont{Kubin}}
  (\bibinfo{publisher}{Elsevier Science}, \bibinfo{year}{2009}),
  vol.~\bibinfo{volume}{15} of \emph{\bibinfo{series}{Dislocations in Solids}},
  chap.~\bibinfo{chapter}{88}, ISBN \bibinfo{isbn}{9780444532855}.

\bibitem[{\citenamefont{{Nabarro}}(1951)}]{Nabarro}
\bibinfo{author}{\bibfnamefont{F.~R.~N.} \bibnamefont{{Nabarro}}},
  \bibinfo{journal}{Proceedings of the Royal Society of London, Series A}
  {\bibinfo{volume}{209}}, \bibinfo{pages}{278} (\bibinfo{year}{1951}).

\bibitem[{\citenamefont{Vitek}(1974)}]{Vitek1974}
\bibinfo{author}{\bibfnamefont{V.}~\bibnamefont{Vitek}},
  \bibinfo{journal}{Crystal Lattice Defects} {\bibinfo{volume}{5}},
  \bibinfo{pages}{pp. 1} (\bibinfo{year}{1974}).

\bibitem[{\citenamefont{Vitek}(2004)}]{Vitek2004}
\bibinfo{author}{\bibfnamefont{V.}~\bibnamefont{Vitek}},
  \bibinfo{journal}{Philosophical Magazine} {\bibinfo{volume}{84}},
  \bibinfo{pages}{415} (\bibinfo{year}{2004}).

\bibitem[{\citenamefont{Dudarev and Derlet}(2005)}]{Dudarev}
\bibinfo{author}{\bibfnamefont{S.~L.} \bibnamefont{Dudarev}} \bibnamefont{and}
  \bibinfo{author}{\bibfnamefont{P.~M.} \bibnamefont{Derlet}},
  \bibinfo{journal}{Journal of Physics: Condensed Matter}
  {\bibinfo{volume}{17}}, \bibinfo{pages}{7097} (\bibinfo{year}{2005}).

\bibitem[{\citenamefont{Mendelev et~al.}(2003)\citenamefont{Mendelev, Han,
  Srolovitz, Ackland, Sun, and Asta}}]{Mendelev}
\bibinfo{author}{\bibfnamefont{M.~I.} \bibnamefont{Mendelev}},
  \bibinfo{author}{\bibfnamefont{S.}~\bibnamefont{Han}},
  \bibinfo{author}{\bibfnamefont{D.~J.} \bibnamefont{Srolovitz}},
  \bibinfo{author}{\bibfnamefont{G.~J.} \bibnamefont{Ackland}},
  \bibinfo{author}{\bibfnamefont{D.~Y.} \bibnamefont{Sun}}, \bibnamefont{and}
  \bibinfo{author}{\bibfnamefont{M.}~\bibnamefont{Asta}},
  \bibinfo{journal}{Philosophical Magazine} {\bibinfo{volume}{83}},
  \bibinfo{pages}{3977} (\bibinfo{year}{2003}).

\bibitem[{\citenamefont{Itakura et~al.}(2012)\citenamefont{Itakura, Kaburaki,
  and Yamaguchi}}]{JAP}
\bibinfo{author}{\bibfnamefont{M.}~\bibnamefont{Itakura}},
  \bibinfo{author}{\bibfnamefont{H.}~\bibnamefont{Kaburaki}}, \bibnamefont{and}
  \bibinfo{author}{\bibfnamefont{M.}~\bibnamefont{Yamaguchi}},
  \bibinfo{journal}{Acta Materialia} {\bibinfo{volume}{60}},
  \bibinfo{pages}{3698 } (\bibinfo{year}{2012}), ISSN
  \bibinfo{issn}{1359-6454}.

\bibitem[{\citenamefont{Ventelon and Willaime}(2010)}]{VentDFT}
\bibinfo{author}{\bibfnamefont{L.}~\bibnamefont{Ventelon}} \bibnamefont{and}
  \bibinfo{author}{\bibfnamefont{F.}~\bibnamefont{Willaime}},
  \bibinfo{journal}{Philosophical Magazine} {\bibinfo{volume}{90}},
  \bibinfo{pages}{1063} (\bibinfo{year}{2010}).

\bibitem[{\citenamefont{Li et~al.}(2004)\citenamefont{Li, Wang, Chang, Cai,
  Bulatov, Ho, and Yip}}]{Cai2004}
\bibinfo{author}{\bibfnamefont{J.}~\bibnamefont{Li}},
  \bibinfo{author}{\bibfnamefont{C.-Z.} \bibnamefont{Wang}},
  \bibinfo{author}{\bibfnamefont{J.-P.} \bibnamefont{Chang}},
  \bibinfo{author}{\bibfnamefont{W.}~\bibnamefont{Cai}},
  \bibinfo{author}{\bibfnamefont{V.~V.} \bibnamefont{Bulatov}},
  \bibinfo{author}{\bibfnamefont{K.-M.} \bibnamefont{Ho}}, \bibnamefont{and}
  \bibinfo{author}{\bibfnamefont{S.}~\bibnamefont{Yip}},
  \bibinfo{journal}{Phys. Rev. B} {\bibinfo{volume}{70}},
  \bibinfo{pages}{104113} (\bibinfo{year}{2004}).

\bibitem[{\citenamefont{Veiga et~al.}(2011)\citenamefont{Veiga, Perez,
  Becquart, Clouet, and Domain}}]{Clouet2011}
\bibinfo{author}{\bibfnamefont{R.}~\bibnamefont{Veiga}},
  \bibinfo{author}{\bibfnamefont{M.}~\bibnamefont{Perez}},
  \bibinfo{author}{\bibfnamefont{C.}~\bibnamefont{Becquart}},
  \bibinfo{author}{\bibfnamefont{E.}~\bibnamefont{Clouet}}, \bibnamefont{and}
  \bibinfo{author}{\bibfnamefont{C.}~\bibnamefont{Domain}},
  \bibinfo{journal}{Acta Materialia} {\bibinfo{volume}{59}},
  \bibinfo{pages}{6963 } (\bibinfo{year}{2011}), ISSN
  \bibinfo{issn}{1359-6454}.

\bibitem[{\citenamefont{Hudson et~al.}(2004)\citenamefont{Hudson, Dudarev, and
  Sutton}}]{Hudson2004}
\bibinfo{author}{\bibfnamefont{T.~S.} \bibnamefont{Hudson}},
  \bibinfo{author}{\bibfnamefont{S.~L.} \bibnamefont{Dudarev}},
  \bibnamefont{and} \bibinfo{author}{\bibfnamefont{A.~P.}
  \bibnamefont{Sutton}}, \bibinfo{journal}{Proceedings of the Royal Society of
  London. Series A: Mathematical, Physical and Engineering Sciences}
  {\bibinfo{volume}{460}}, \bibinfo{pages}{2457} (\bibinfo{year}{2004}).

\bibitem[{\citenamefont{Kumar et~al.}(2012)\citenamefont{Kumar, Durgaprasad,
  Dutta, and Dey}}]{Kumar2012}
\bibinfo{author}{\bibfnamefont{N.~N.} \bibnamefont{Kumar}},
  \bibinfo{author}{\bibfnamefont{P.}~\bibnamefont{Durgaprasad}},
  \bibinfo{author}{\bibfnamefont{B.}~\bibnamefont{Dutta}}, \bibnamefont{and}
  \bibinfo{author}{\bibfnamefont{G.}~\bibnamefont{Dey}},
  \bibinfo{journal}{Computational Materials Science}
  {\bibinfo{volume}{53}}, \bibinfo{pages}{258 } (\bibinfo{year}{2012}),
  ISSN \bibinfo{issn}{0927-0256}.

\bibitem[{\citenamefont{Hudson et~al.}(2005)\citenamefont{Hudson, Dudarev,
  Caturla, and Sutton}}]{Hudson2005}
\bibinfo{author}{\bibfnamefont{T.~S.} \bibnamefont{Hudson}},
  \bibinfo{author}{\bibfnamefont{S.~L.} \bibnamefont{Dudarev}},
  \bibinfo{author}{\bibfnamefont{M.~J.} \bibnamefont{Caturla}},
  \bibnamefont{and} \bibinfo{author}{\bibfnamefont{A.~P.}
  \bibnamefont{Sutton}}, \bibinfo{journal}{Philosophical Magazine}
  {\bibinfo{volume}{85}}, \bibinfo{pages}{661} (\bibinfo{year}{2005}).

\bibitem[{\citenamefont{Kosevich}(2006)}]{Kosevich2006}
\bibinfo{author}{\bibfnamefont{A.}~\bibnamefont{Kosevich}},
  \emph{\bibinfo{title}{The Crystal Lattice: Phonons, Solitons, Dislocations,
  Superlattices}} (\bibinfo{publisher}{John Wiley \& Sons},
  \bibinfo{year}{2006}), ISBN \bibinfo{isbn}{9783527606931}.

\bibitem[{\citenamefont{Jo\'os and Duesbery}(1997)}]{Joos1997}
\bibinfo{author}{\bibfnamefont{B.}~\bibnamefont{Jo\'os}} \bibnamefont{and}
  \bibinfo{author}{\bibfnamefont{M.~S.} \bibnamefont{Duesbery}},
  \bibinfo{journal}{Phys. Rev. B} {\bibinfo{volume}{55}},
  \bibinfo{pages}{11161} (\bibinfo{year}{1997}).

\bibitem[{\citenamefont{Granato and Lucke}(1956)}]{Granato1956}
\bibinfo{author}{\bibfnamefont{A.}~\bibnamefont{Granato}} \bibnamefont{and}
  \bibinfo{author}{\bibfnamefont{K.}~\bibnamefont{Lucke}},
  \bibinfo{journal}{Journal of Applied Physics} {\bibinfo{volume}{27}},
  \bibinfo{pages}{583} (\bibinfo{year}{1956}).

\bibitem[{\citenamefont{Landau and Lifshits}(1975)}]{landau1975classical}
\bibinfo{author}{\bibfnamefont{L.}~\bibnamefont{Landau}} \bibnamefont{and}
  \bibinfo{author}{\bibfnamefont{E.}~\bibnamefont{Lifshits}},
  \emph{\bibinfo{title}{The classical theory of fields}},
  vol.~\bibinfo{volume}{2} (\bibinfo{publisher}{Butterworth-Heinemann},
  \bibinfo{year}{1975}).

\bibitem[{\citenamefont{Zwanzig}(2001)}]{Zwanzig}
\bibinfo{author}{\bibfnamefont{R.}~\bibnamefont{Zwanzig}},
  \emph{\bibinfo{title}{Nonequilibrium Statistical Mechanics}}
  (\bibinfo{publisher}{Oxford University Press}, \bibinfo{year}{2001}), ISBN
  \bibinfo{isbn}{9780195140187}.

\bibitem[{\citenamefont{Reif}(2008)}]{reif}
\bibinfo{author}{\bibfnamefont{F.}~\bibnamefont{Reif}},
  \emph{\bibinfo{title}{Fundamentals of Statistical and Thermal Physics}},
  McGraw-Hill series in fundamentals of physics (\bibinfo{publisher}{Waveland
  Press}, \bibinfo{year}{2008}), ISBN \bibinfo{isbn}{9781577666127}.

\bibitem[{\citenamefont{Saito and Matsumoto}(2008)}]{dSFMT}
\bibinfo{author}{\bibfnamefont{M.}~\bibnamefont{Saito}} \bibnamefont{and}
  \bibinfo{author}{\bibfnamefont{M.}~\bibnamefont{Matsumoto}}, in
  \emph{\bibinfo{booktitle}{Monte Carlo and Quasi-Monte Carlo Methods 2006}},
  edited by \bibinfo{editor}{\bibfnamefont{A.}~\bibnamefont{Keller}},
  \bibinfo{editor}{\bibfnamefont{S.}~\bibnamefont{Heinrich}}, \bibnamefont{and}
  \bibinfo{editor}{\bibfnamefont{H.}~\bibnamefont{Niederreiter}}
  (\bibinfo{publisher}{Springer Berlin Heidelberg}, \bibinfo{year}{2008}), pp.
  \bibinfo{pages}{607--622}.

\bibitem[{\citenamefont{Marchaj}(2002)}]{sail}
\bibinfo{author}{\bibfnamefont{C.}~\bibnamefont{Marchaj}},
  \emph{\bibinfo{title}{Sail performance: techniques to maximize sail power}}
  (\bibinfo{publisher}{International Marine/McGraw-Hill},
  \bibinfo{year}{2002}), ISBN \bibinfo{isbn}{9780071413107}.

\bibitem[{\citenamefont{Caillard}(2010)}]{Caillard2}
\bibinfo{author}{\bibfnamefont{D.}~\bibnamefont{Caillard}},
\emph{\bibinfo{title}{Kinetics of dislocations in pure Fe. Part I. In situ straining experiments at room temperature}}, \bibinfo{journal}{Acta Materialia} {\bibinfo{volume}{58}},
  \bibinfo{pages}{3493 - 3503} (\bibinfo{year}{2010}).


\end{thebibliography}
\end{document}